\documentclass[%
 reprint,
 amsmath,amssymb,
 aps,
]{revtex4-2}

\usepackage{graphicx}
\usepackage{subcaption}
\usepackage{dcolumn}
\usepackage{bm}
\usepackage{braket}
\usepackage{amsmath}
\usepackage{enumitem}

\begin{document}

\preprint{APS/123-QED}

\title{Many-body effects in the line radiative transfer equation}
\author{Boy Lankhaar}
 \email{boy.lankhaar@astro.uio.no}
\affiliation{%
Institute of Theoretical Astrophysics, University of Oslo, PO Box 1029, Blindern 0315, Oslo, Norway
}%
\affiliation{%
Department of Space, Earth and Environment, Chalmers University of Technology, Onsala Space Observatory, 439 92 Onsala, Sweden
}%

\date{\today}

\begin{abstract}
The radiative transfer equation for spectral lines from an extended gas is derived from first principles, treating the gas as a system of many atoms/molecules rather than isolated ones. Line broadening effects are assumed to be dominated by particle motions (Doppler effect), but collisional broadening effects are included in the impact approximation. We retrieve the canonical radiative transfer equation under the condition that the optical depth over a coherence length, defined as the transition-levels lifetime times the speed of light, is much lower than unity. For other cases, the line radiative transfer equation contains a correction factor whose magnitude depends exponentially on a quantity that we call the coherent optical depth. We compute that many-body effects affect line radiative transfer of strongly emitting and astronomically ubiquitous radio- and submillimeter lines, such as the HI 21 cm line, and rotational transitions of the main isotopologue of CO. These results imply that care must be taken when interpreting observations relying on these lines, as many-body effects can significantly alter emergent line profiles and bias inferred physical conditions of the emitting gas. Finally, we propose a simple laboratory experiment that would reveal many-body effects in the transfer of radiation, which could furthermore offer a cost-effective means of constraining fundamental molecular parameters.
\end{abstract}

\maketitle
%
%
\section{Introduction}
The line radiative transfer equation provides the fundamental framework for describing how radiation interacts with matter across a wide range of environments. It underpins the treatment of astrophysical line radiation transport in both the interstellar medium \cite{rybicki:08} and stellar atmospheres \cite{mihalas:78}, and is central in atmospheric physics \cite{janssen:94} and plasma spectroscopy \cite{griem:12}. Also in phenomena such as radiative trapping, which affect a wide array of experimental situations, the radiative transfer equation is implicit \cite{molisch:98}. In its canonical form, the radiative transfer equation reads
\begin{align}
\label{eq:lrt_can}
\frac{d I_{\nu}}{ds} = - \kappa_{\nu} I_{\nu} + \epsilon_{\nu},
\end{align}
and describes the change in radiation intensity $I_{\nu}$ along a path $ds$ as the sum of absorption and stimulated emission processes, represented by term $\kappa_{\nu} I_{\nu}$, with $\kappa_{\nu}$ the opacity coefficient, and spontaneous emission processes, represented by the emissivity term $\epsilon_{\nu}$.

The foundations of the radiative transfer equation can be traced to the early work of \citet{einstein:17}, who introduced the basic processes of radiation-matter interaction---absorption, and stimulated and spontaneous emission---and to \citet{dirac:27}, who rigorously quantified these processes from a full quantum mechanical theory. However, these works did not explicitly relate these processes to radiation transport. Instead, the first quantum-mechanical derivations connecting radiation–matter interaction to the transport of radiation, leading to equations similar to Eq.~(\ref{eq:lrt_can}), were given by Refs.~\cite{landi:72,machacek:78,machacek:79}, and were later refined in \citet{landi:83} to also include polarization in both the radiation and atomic states. Combined with the statistical equilibrium equations \cite{cooper:82,landi:83}, this formalism is the foundation of radiative transfer theory.

%
The radiative transfer equation of Eq.~(\ref{eq:lrt_can}) is a Boltzmann-like transport equation, that tacitly assumes that photons behave as particles or rays with a well defined position and momentum. However, formally, a proper quantum treatment must account for the uncertainty between photon position and momentum \cite{newton:49,brittin:62,wolf:76,sudarshan:81}. Such a quantum mechanical transport theory models instead the transport of a Wigner quasiprobability distribution. Derivations that leverage this approach confirm that Eq.~(\ref{eq:lrt_can}) is valid for broad-band, slowly varying fields \cite{cooper:84}. More recent work \cite{rosato:10,rosato:11,rosato:13} has shown that the Boltzmann-like equation holds, provided the photon mean free path $\ell_{\nu}^{\mathrm{mfp}}=\kappa_{\nu}^{-1}$ is much larger than the radiation coherence length $\ell_{\mathrm{r.c.}}\sim c/\Delta \omega_{1/2}$, where $\Delta \omega_{1/2}$ is the half-width of the line profile. In the case $\ell_{\nu}^{\mathrm{mfp}}/\ell_{\mathrm{r.c.}}\lesssim 1$, photon position-momentum uncertainty reduce the effective opacity and enhances photon escape \cite{rosato:11}.

Another challenge to the validity of Eq.~(\ref{eq:lrt_can}) arises from its assumption of Gaussian photon statistics, which implies uncorrelated photon modes \cite{landi:83}. This assumption breaks down in the strong-field regime of lasers or masers, when stimulated emission rates exceed the lifetime of transition states. For these intense radiation fields, matter-radiation interactions will build correlations between different modes of the radiation field \cite{menegozzi:78, wyenberg:21}. The radiative transfer formalism of Eq.~(\ref{eq:lrt_can}) must then be replaced by the Maxwell–Bloch equations \cite{lamb:64}, a time-dependent description of the coupled evolution of the radiation field, and the the atomic or molecular polarization and populations. Unlike the radiative transfer equation, this framework explicitly accounts for coherence between radiation modes and the time-dependent dynamics of the transition states and polarization, as required in the non-steady-state laser regime \cite{sargent:74}.

A related non–steady-state phenomenon is superradiance, predicted by \citet{dicke:54} as an emission mode arising when atoms or molecules in a gas form a correlated state in which their dipoles align, leading to an emissivity that scales with the square of the number of radiators. In Dicke's formalism of radiation-matter interactions, the interaction of a radiation field and a cloud of gas containing many particles, is explicitly modeled as a many-body problem, thus allowing for different atoms or molecules to get correlated. For extended gases with dimensions much larger than the radiation wavelength, particles become correlated along the forward direction of the dominant radiation mode \cite{dicke:54,dicke:64}. Superradiance was first observed experimentally by \citet{skribanowitz:73}, who showed that the superradiant pulse characteristics agreed well with theoretical predictions. These predictions were derived from the Maxwell–Bloch equations, that were shown to be formally equivalent to Dicke’s original formalism \cite{arecchi:70,lehmberg:70,gross:82}. \citet{skribanowitz:73} also found that inhomogeneous broadening had only a marginal effect. Superradiance exemplifies a regime where collective interactions between emitters fundamentally alter the character of radiation–matter coupling.

In this light, it is noteworthy that derivations of the radiative transfer equation have generally considered the interaction between a radiation field and a cloud of gas, not as an explicit many-body problem, but instead as the incoherent sum of the interactions between isolated atoms or molecules, and the radiation field \cite{landi:72, landi:83, machacek:78}. Within this framework, interparticle correlations are explicitly excluded. A fully many-body formalism, that does allow for interparticle correlations, is more general and should, in principle, reduce to the radiative transfer equation under appropriate limits. To the best of our knowledge, no derivation has explicitly verified the absence of interparticle correlations that justifies this central assumption underlying the radiative transfer equation, i.e.~that it can be represented as the incoherent sum of independent interactions between the radiation field and isolated particles. Here, we address this issue and rigorously derive the radiative transfer equation, starting from the many-body radiation-matter interaction framework of \citet{dicke:54}. We assume that photons are localized in position-momentum space, fulfilled when $\ell_{\nu}^{\mathrm{mfp}}\gg \ell_{\mathrm{r.c.}}$ \cite{rosato:11}. We explicitly model line broadening effects due to Doppler effects \cite{fermi:32}, and we model pressure broadening, including the impact of collisions on transition level lifetimes following the impact theory of \citet{baranger:58}. Specializing to weak (non-laser) radiation fields, and the case of a Doppler broadened line, we find that the canonical line radiative transfer equation of Eq.~(\ref{eq:lrt_can}), is in fact a special case of the many-body line radiative transfer equation,
\begin{align}
\label{eq:lrt_mb}
\frac{d I_{\nu}}{ds} = e^{-\tau_{\nu}^{\mathrm{coh}}}\left[ - \kappa_{\nu} I_{\nu} + \epsilon_{\nu} \right],
\end{align}
in the limit of low coherent optical depth, $\tau_{\nu}^{\mathrm{coh}} = \kappa_{\nu} \ell_{\mathrm{coh}}$, where the coherence length, $\ell_{\mathrm{coh}} = c \tau_M$, is the product of the speed of light, $c$, and the lifetime of the transition levels, $\tau_M$. For the vast majority of naturally occurring spectral lines, indeed $\tau_{\nu}^{\mathrm{coh}} \ll 1$, and Eq.~(\ref{eq:lrt_can}) is an accurate description of the radiative transfer, yet in specific parameter regimes---that can be produced experimentally, and naturally occur for important molecular and atomic lines in astronomy---many-body effects are expected to impact the radiative transfer and Eq.~(\ref{eq:lrt_mb}) has to be preferred.

This paper is built up as follows. In Sec.~II, a general form of the many-body radiative transfer equation is derived from first principles. We model the interaction between a multimode radiation field and a collection of moving molecules using Dicke’s Hamiltonian formalism, and include collisional decorrelation effects via a density matrix approach and the impact approximation. In Sec.~III, we examine a form of the radiative transfer equation, focusing on the regime where many-body effects dominate correction elements. We show that we recover the canonical radiative transfer equation, as a limiting case of the general many-body formalism, valid when the coherent optical depth is much less than unity. In other cases, the many-body radiative transfer equation has to be preferred. In Sec.~IV, we  provide a narrative summary of the derivation, highlighting the assumptions and physical interpretation of the many-body radiative transfer equation. Afterwards, we compare our formalism to previous quantum-mechanical derivations of the radiative transfer equation. General analytic solutions to the many-body radiative transfer equation are derived, illustrating how collective effects modify emergent line profiles. Finally, we discuss the physical meaning and scaling behavior of the coherent optical depth, propose a laboratory experiment capable of detecting the predicted suppression of absorption due to many-body effects, and provide order-of-magnitude estimates showing that many-body effects significantly alter the propagation of radiation in astrophysical lines such as H I and CO. We conclude in Sec.~V.

\section{A general form of the many-body radiative transfer equation}
In the following, we derive a general form of the radiative transfer equation for spectral lines by treating the gas as a many-body system, rather than as an ensemble of isolated particles. Our starting point is a Hamiltonian formalism that captures the interaction between a multimode radiation field and a collection of moving two-level molecules or atoms. In subsection A, we introduce Dicke’s many-body Hamiltonian that includes internal molecular states, kinetic energy from translational motion, the quantized radiation field, and the dipole-mediated matter-radiation interaction. In subsection B, we incorporate collisional effects, which we assume is the dominant mechanism of decorrelation, by coupling the molecular ensemble to a bath using density matrix theory and the impact approximation. Finally, in subsection C, we derive a general expression for the time evolution of the radiation field interacting with a molecular ensemble. This formulation lays the groundwork for analyzing both canonical and corrected forms of the line radiative transfer equation in the subsequent section.

\begin{table*}[h]
  \caption{List of symbols and operators used in the manuscript.}
  \begin{ruledtabular}
  \begin{tabular}{l l l}
  Symbol & Units & Meaning / definition \\ \hline
  $I_\nu$ & $\mathrm{W\,m^{-2}\,Hz^{-1}\,sr^{-1}}$ & Specific intensity of radiation at frequency $\nu$. \\
  $S_\nu$ & $\mathrm{W\,m^{-2}\,Hz^{-1}\,sr^{-1}}$ & Source function $S_\nu\equiv \epsilon_\nu/\kappa_\nu$. \\
  $\kappa_\nu$ & $\mathrm{m^{-1}}$ & Absorption coefficient (line opacity); definition Eq.~(\ref{eq:rad_trans_0}). \\
  $\epsilon_\nu$ & $\mathrm{W\,m^{-3}\,Hz^{-1}\,sr^{-1}}$ & Emissivity; definition Eq.~(\ref{eq:rad_trans_0}). \\
  $\phi_\nu$ & $\mathrm{Hz^{-1}}$ & Line profile; for Doppler $\phi_\nu=\frac{1}{\sqrt{\pi}\,\Delta\nu}\exp[-(\nu-\nu_0)^2/\Delta\nu^2]$. \\
  $\phi'_\nu$ & $\mathrm{Hz^{-1}}$ & Many-body adjusted profile $\phi'_\nu = e^{-\tau^{\mathrm{coh}}_\nu}\phi_\nu$. \\
  $\nu_0,\ \omega_0$ & Hz, s$^{-1}$ & Line-center frequency and angular frequency ($\omega_0=2\pi\nu_0=2\pi c / \lambda$). \\
  $\Delta\nu$ & Hz & Line width in frequency units. \\
  $\tau_\nu$ & -- & Optical depth. \\
  $\tau_{\nu}^{\mathrm{coh}}$ & -- & Coherent optical depth ($\tau_{\nu}^{\mathrm{coh}} = \kappa_{\nu} \ell_{\mathrm{coh}}$). \\
  $\ell_{\rm coh}$ & m & Coherence length of the medium, $\ell_{\rm coh}=c\,\tau_M$). \\
  $\ell_{D},\ \ell_{\mathrm{r.c.}}$ & m & Doppler coherence length ($c \tau_D$) and radiative coherence length $c/\Delta \nu$. \\
  $\tau_D,\ \tau_L,\ \tau_M$ & s & Doppler, elastic collision, and lifetime decorrelation timescales. \\
  $A_0$ & s$^{-1}$ & Einstein A coefficient for spontaneous emission. For transition $u \to l$, $A_0 = g_u A_{ul}$. \\
  $B_0$ & m$^3$J$^{-1}$s$^{-2}$ & Einstein B coefficient for stimulated processes. For transition $u \to l$, $B_0 = g_uB_{ul}=g_lB_{lu}$. \\
  $\bar{n}$ & -- & Mean photon occupation number; $B_0\bar i = A_0 \bar n$ with $\bar i\equiv\!\int\phi_\nu \frac{d\Omega}{4\pi}I_\nu\,d\nu$. \\
  $N$ & -- & Number of molecules in the ensemble. \\
  $V$ & m$^{3}$ & Quantization volume. \\
  $\mathcal{N}$ & m$^{-3}$ & Number density, $N/V$, of molecules in the ensemble. \\
  $n_\pm$ & -- & Level populations (upper $+$, lower $-$) per molecule ($n_+ = n_u/g_u$ and $n_-=n_l/g_l$). \\
  $x_{\rm ab}$ & -- & Abundance fraction of the species. \\
  $Q$ & -- & Partition function. \\
  $b$ & -- & Non-thermal (turbulent) broadening factor over $\ell_{\rm coh}$, entering $\phi_\nu$  \\
  $\lambda$ & m & Wavelength ($\lambda=2\pi c/\omega_0$). \\
  $c,\ h,\ \hbar,\ k_B$ & SI & Physical constants (speed of light, Planck, reduced Planck, Boltzmann). \\
  $\mu$ & C\,m (or D) & Permanent dipole moment of the molecule (used in rotational estimates). \\
  $B_{\rm rot}$ & Hz & Rotational constant for a linear rotor. \\
  $J$ & -- & Rotational quantum number (upper level in $J\!\to\!J\!-\!1$). \\
  $g_u,\ g_l$ & -- & Statistical weights of upper/lower levels. \\
  $m,\ M$ & kg & Reduced mass for collisions; molecular mass for translation. \\
  $\sigma$ & m$^2$ & Effective collisional cross-section. \\
  $\hat{\boldsymbol{r}}_j,\ \hat{\boldsymbol{P}}_j,\ \hat{\boldsymbol{v}}_j$ & m, kg\,m\,s$^{-1}$, m\,s$^{-1}$ & Position, momentum, velocity operators of molecule $j$. \\
  $\hat H_M$ & J & Molecular two-level Hamiltonian ($|+\rangle,|-\rangle$ transition). \\
  $\hat H_K$ & J & Translational kinetic Hamiltonian; $\hat H_K=\sum_j \hat{\mathbf P}_j^2/2M$. \\
  $\hat H_R$ & J & Radiation field Hamiltonian; $\hat H_R=\sum_k \hbar\omega_k(\hat n_k+\tfrac12)$. \\
  $\hat V$ & J & Dipole interaction Hamiltonian (RWA/dipole approx.). \\
  $\hat H_B,\ \hat W$ & J & Bath Hamiltonian and the interaction operation describing interaction with transition molecules. \\
  $\hat \rho(t),\ \hat \rho_I(t),\ \rho_0$ & -- & Total density operator, its interaction-picture form and value at $t=0$. \\
  $\hat a_{\boldsymbol{k}},\ \hat a_{\boldsymbol{k}}^\dagger$ & -- & Photon annihilation/creation operators of mode $\boldsymbol{k}$. \\
  $\hat n_{\boldsymbol{k}}$ & -- & Photon number operator ($\hat n_{\boldsymbol{k}}=\hat a_{\boldsymbol{k}}^\dagger \hat a_{\boldsymbol{k}}$). \\
  $\hat R_{j\pm},\ \hat R_{j3}$ & -- & Two-level (pseudo-spin) ladder and projection operators for molecule $j$; $[\,\hat R_{j+},\hat R_{j-}\,]=2\hat R_{j3}$. \\
  $\tilde R_{j\pm}(t)$ & -- & Interaction-picture state operators with bath-induced decorrelation. \\
  $\hat U(t)$ & -- & Time-evolution operator for bath coupling (used in $\tilde R_{j\pm}$). \\
  $\hat A(t)$ & -- & Non-unitary averaged propagator after tracing out the bath. \\
  $\hat{\mathcal{T}}\ (\tilde{\mathcal{T}})$ & -- & Time (anti)ordering operator. \\
  $\hat B(t)$ & -- & Radiation–matter coupling block (so that $\hat V(t)=\hat B(t)+\hat B^\dagger(t)$). \\
  $\hat i_k$ & -- & Intensity operator for mode $\boldsymbol{k}$. \\
  $\hat\sigma(t,t'),\ \hat\gamma(t,t')$ & -- & Stimulated and spontaneous radiative operators. \\
  $g_k$ & s$^{-1}$ & Light–matter coupling strength ($g_k=\sqrt{\omega_k/(2\hbar\epsilon_0 V)}\,\mathbf e_k\!\cdot\!\hat{\mathbf d}$). \\
  $\hat{\boldsymbol{ d}}$ & C\,m & Molecular dipole operator. \\
  $t,t',t'',\cdots$ & s & time parameters for the system evolution, with $0\le t''' \le t'' \le t' \le t$.\\
  $\tau,\tau',\tau'',\cdots$ & s & (positive) time integration parameters, with $\tau = t-t',\ \tau' = t'-t'', \cdots$.\\
  $i\equiv\sqrt{-1}$ & -- & Imaginary unit; used in phases and commutators. \\
  \end{tabular}
  \end{ruledtabular}
\end{table*}

\subsection{Dicke's many-body Hamiltonian}
We consider the interaction between a multimode radiation field and a collection of $N$ molecules \footnote{For brevity, we use the term ``molecules'' throughout; the formalism applies equally to atoms.}, each possessing a pair of internal energy eigenstates relevant to the radiative transition under consideration, denoted by $+$ or $-$, and separated in energy by $\hbar \omega_0 = h \nu_0 = hc / \lambda$. While the molecular eigenstates are in general rovibronic, the formalism is general and applies to rotational, vibrational, or electronic transitions. The molecules are distributed in a volume $V=4\pi L^3/3$ of dimension $L$, which is much larger than the typical wavelength of the radiation field under consideration. The incoming radiation field is assumed to behave according to Gaussian statistics, while the molecules are distributed randomly at positions $\boldsymbol{r}_j$, where $j$ indicates a molecule in the volume. We focus on transition lines that are broadened primarily through Doppler broadening. We describe the interaction between the radiation field and the molecules, following a similar approach as \citet{dicke:54}. We let $\hat{R}_{j\pm}$ be the operator that promotes/demotes molecule, $j$, from the $\mp$ to the $\pm$ state, while $\hat{R}_{j3}$ is a projection-like operator that returns $\pm \frac{1}{2}$ when molecule $j$ is in the $\pm$ state. The operators $\hat{R}_{j \pm}$ and $\hat{R}_{j3}$ are part of a Lie-algebra. Their commutation relations read $[\hat{R}_{j+},\hat{R}_{j-}]=2\hat{R}_{j3}$ and $[\hat{R}_{j3},\hat{R}_{j\pm}] = \pm \hat{R}_{j\pm}$, where we furthermore note that commutators between different particles return $0$. 

With $\hbar \omega_0$ as the difference in energy between states $+$ and $-$, the molecular internal energy-Hamiltonian reads: 
\begin{subequations}
\begin{align}
\hat{H}_M = \hbar \omega_0\sum_j \hat{R}_{j3}.
\end{align}
Besides having an internal energy, molecules are also endowed with a linear momentum that is due to their motion in space: 
\begin{align}
\hat{H}_K = \sum_j \frac{\hat{P}_j^2}{ 2M},
\end{align}
where $\hat{\boldsymbol{P}}_j=-i\hbar d/d\boldsymbol{r}_j$ is the momentum operator of particle $j$, and $M$ is the molecular mass. 

The radiation field is described while neglecting the polarization. We describe the radiation field in second quantization in the usual way. We let $\hat{a}_{\boldsymbol{k}}$ ($\hat{a}_{\boldsymbol{k}}^{\dagger}$) be the annihilation (creation) operator of photon-mode $\boldsymbol{k}$, that has a direction $\hat{k}$ and associated frequency $\omega_k$. With the associated energy of photons $\hbar \omega_k$, the Hamiltonian of the radiation field is 
\begin{align}
\hat{H}_R &= \sum_{\boldsymbol{k}}\hbar \omega_k (\hat{a}_{\boldsymbol{k}}^{\dagger} \hat{a}_{\boldsymbol{k}} + \frac{1}{2}) \nonumber \\ 
&= \sum_{\boldsymbol{k}}\hbar \omega_k (\hat{n}_{\boldsymbol{k}}+ \frac{1}{2}),
\end{align}
\end{subequations}
where we defined the photon-number operator $\hat{n}_{\boldsymbol{k}}$.

Radiation of wavelength close to, $\lambda = 2\pi c / \omega_0$, may induce a transition between both states of the molecule, through interaction with the transition-dipole moment. The interaction Hamiltonian, $\hat{V}$, that describes the interaction between an oscillating electric field and the molecular ensemble, in the dipole-approximation, is \cite{dicke:54, houde:13, houde:22}
\begin{align}
\hat{V} = -i \sum_{\boldsymbol{k}} \left[\hbar g_{\boldsymbol{k}} \hat{a}_{\boldsymbol{k}} \sum_j \hat{R}_{j +} e^{i \boldsymbol{k}\cdot \hat{\boldsymbol{r}}_j} - \hbar g_{\boldsymbol{k}}\hat{a}_{\boldsymbol{k}}^{\dagger} \sum_j \hat{R}_{j -} e^{-i \boldsymbol{k}\cdot \hat{\boldsymbol{r}}_j} \right],
\end{align}
where the summations run over all photon modes, $\boldsymbol{k}$, and particles, $j$. The interaction constant, that is in frequency units, reads
\begin{align}
g_{\boldsymbol{k}} = \sqrt{\frac{\omega_k}{2 \hbar \epsilon_0 V}} \boldsymbol{e}_{\boldsymbol{k}} \cdot \hat{\boldsymbol{d}}, \nonumber
\end{align} 
where $\epsilon_0$ is the electric constant, $\hat{\boldsymbol{d}}$ is the dipole-operator, $\boldsymbol{e}_{\boldsymbol{k}}$ is the direction of the electric field.

It may be regarded advantageous to define the operators $\hat{R}_{\boldsymbol{k} \pm} = \sum_j \hat{R}_{j\pm} e^{\pm i \boldsymbol{k} \cdot \hat{\boldsymbol{r}}_j}$ and $\hat{R}_{3} = \sum_j \hat{R}_{j3}$, that allow for simultaneous eigenfunctions as $\hat{R}_{\boldsymbol{k}}^2 = \hat{R}_{\boldsymbol{k} +}^2 + \hat{R}_{\boldsymbol{k} -}^2 + \hat{R}_{3}^2$ commutes with both $\hat{H}_M$ and $\hat{V}$ \cite{dicke:54}. However, we explicitly refrain from using these collective operators for our Hamiltonians, as they do not commute with $\hat{H}_K$: the global operators disintegrate due to the motion of the individual molecules. 

As a final note, we emphasize that we model the interaction as an effective two-level system, with states denoted by $\pm$. In practice, the upper and lower levels of a spectral transition are generally degenerate. We account for this degeneracy post hoc by interpreting $n_{\pm}$ as populations per degenerate sublevel, so that the total upper/lower level populations are $n_{u/l}=g_{u}\,n_{\pm}$, where $g_{u/l}$ are the corresponding degeneracies. This corresponds to assuming that all degenerate sublevels are populated equally and that no coherences between them are present. Such an assumption is appropriate when the pumping and relaxation processes do not induce atomic/molecular polarization. If the medium interacts with strong polarized and/or anisotropic radiation, one must generally relax this assumption and instead treat the sublevel density matrix and solve the corresponding polarized radiative-transfer and statistical-equilibrium problem \cite{landi:06,bommier:97,lankhaar:20a}. This is beyond the scope of the current paper.

\subsection{Density matrix theory and collisional decorrelation}
The Hamiltonians $\hat{H}_M + \hat{H}_K + \hat{H}_R+\hat{V}$ describe a closed system of a molecular ensemble, interacting with a multimode radiation field through the interaction Hamiltonian $\hat{V}$. Apart from the matter-radiation interaction, there may be additional interactions between the molecular ensemble, or radiation field, with other particles, that we refer to as the bath. Interactions may occur between a bath and the radiation field operators, for instance, through absorption or emission close to the transition frequency by a different gas phase. Alternatively, interactions may occur between a bath and the position operators, appearing in the terms $e^{i \boldsymbol{k}\cdot \hat{\boldsymbol{r}}_{j}}$. Third, interactions between a bath and the molecular state operators can be present. Here, we assume that dominant decorrelation pertains to the molecular state operators. We assume that the molecular ensemble is actively interacting with a bath, which we take to be all other molecules in the gas, through collisions. Collisional interactions will degrade molecular state operators $\hat{R}_{j\pm}$. 

We extend the Hamiltonians of the Dicke system with Hamiltonians that pertain to the bath, $\hat{H}_B$, and its interaction term with the molecular operators, $\hat{W}$: $$\hat{H} = \hat{H}_M + \hat{H}_K + \hat{H}_R+\hat{V} + \hat{H}_P + \hat{W}.$$ Now, we proceed to consider the total density operator $\hat{\rho}(t)$, which includes both the Dicke system and the bath. We note the Schr\"{o}dinger equation,
\begin{subequations}
\begin{align}
\label{eq:dens_general}
\frac{d \hat{\rho}(t)}{d t} = -i \hbar^{-1} \left[\hat{H},\hat{\rho}(t) \right],
\end{align}
that describes the time-evolution of the density-operator. If we define the density-operator in the interaction picture $\hat{\rho}_I (t) = e^{i[\hat{H}_M + \hat{H}_K + \hat{H}_R +  \hat{H}_B + \hat{W}]} \hat{\rho}(t) e^{-i[\hat{H}_M + \hat{H}_K + \hat{H}_R +  \hat{H}_B + \hat{W}]}$, then it is easy to recognize that its time-dependence, is
\begin{align}
\label{eq:dens_interaction}
\frac{d \hat{\rho}_I(t)}{d t} = -i \hbar^{-1} \left[\hat{V}(t),\hat{\rho}_I(t) \right],
\end{align}
where $$\hat{V}(t) = e^{i[\hat{H}_M + \hat{H}_K + \hat{H}_R +  \hat{H}_B + \hat{W}]t} \hat{V} e^{-i[\hat{H}_M + \hat{H}_K + \hat{H}_R +  \hat{H}_B + \hat{W}]t}.$$ 
We note the exact relation of the density-operator, in the interaction picture, to the density-operator at $t=0$, $\hat{\rho}_0$, by
\begin{align}
\label{eq:densI_exact}
\hat{\rho}_I(t) = \hat{\mathcal{T}}e^{-i \hbar^{-1} \int_0^t dt' \ \hat{V}(t')} \hat{\rho}_0 \tilde{\mathcal{T}}e^{i \hbar^{-1} \int_0^t dt' \ \hat{V}(t')} ,
\end{align}
\end{subequations}
where $\hat{\mathcal{T}}$ ($\tilde{\mathcal{T}}$) denotes the (anti-)time ordering operator. We now evaluate the interaction operator in the interaction picture, that includes decorrelation due to collisions with the bath \cite{baranger:58, fano:57, fiutak:62, omont:72},
\begin{align}
\hat{V}(t) &= 
-i \hbar \sum_{\boldsymbol{k}}g_{\boldsymbol{k}} \left[ \hat{a}_{\boldsymbol{k}} \sum_j \tilde{R}_{j +} (t) e^{i \boldsymbol{k}\cdot \hat{\boldsymbol{r}}_j} e^{i (\omega_0 - \omega_k)t} e^{i \boldsymbol{k} \cdot \boldsymbol{v}_jt} \right. \nonumber \\ &- \left. \hat{a}_{\boldsymbol{k}}^{\dagger} \sum_j \tilde{R}_{j -}(t) e^{-i \boldsymbol{k}\cdot \hat{\boldsymbol{r}}_j} e^{-i (\omega_0 - \omega_k)t} e^{-i \boldsymbol{k} \cdot \hat{\boldsymbol{v}}_jt} \right] \nonumber \\
&= \hat{B} (t) + \hat{B}^{\dagger} (t),
\end{align}
where, besides adding phase factors due to the energy of the associated particles, we have transformed the operators $\tilde{R}_{j\pm}(t) = \hat{U}^{\dagger}(t) \hat{R}_{j\pm} \hat{U}(t)$, where $\hat{U}(t) = e^{-i(\hat{H}_B + \hat{W})t/\hbar}$. The phase factors $e^{i \boldsymbol{k} \cdot \hat{\boldsymbol{v}}_jt}$, that follow from the non-commutativity of $\hat{H}_K$ and the exponential $e^{i \boldsymbol{k}\cdot \hat{\boldsymbol{r}}_j}$, are functions of the particle velocities $\hat{\boldsymbol{v}}_j=\hat{\boldsymbol{P}}_j/M$ and represent the Doppler shift that will eventually lead to a decorrelation function due to the random motions of the particles: 
\begin{align}
\label{eq:doppler_dec}
\mathrm{tr}\left\{e^{i \boldsymbol{k} \cdot \hat{\boldsymbol{v}}_jt} \hat{\rho}_0^{(K)}\right\} \to e^{-(t/\tau_D)^2},
\end{align} 
with $\tau_D = [\lambda / 2\pi] \sqrt{2M/k_B T } $, where $M$ is the particle mass and $T$ the kinetic temperature. Finally, the decomposition of $\hat{V}$, into the $\hat{B} (t)$ operators will be advantageous when analyzing commutator algebra.

Evaluating the matrix-elements of the $\tilde{R}_{j\pm}(t)$ operators, and products of them, are a known problem in line broadening theory\cite{baranger:58, fano:57, fiutak:62, omont:72}. In Appendix~A, we consider this problem in detail, for an arbitrary order of operator-products that emerge in the many-body expansion of the radiative transfer equation. Here, we recall some key elements of the approximations that will be used. We modeled the $\hat{R}_{j\pm}$ operators as being affected by a bath of (other) gaseous molecules, that interact with the particles, which the $\hat{R}_{j\pm}$ operators pertain to. Thus in effect, the propagation $\hat{U}^{\dagger}(t) \hat{R}_{j\pm} \hat{U}(t)$ represents the effects of scattering events on the operator $\hat{R}_{j\pm}$. In modeling these scattering events, we follow \citet{baranger:58}, and leverage the impact approximation, where subsequent scattering events are modeled as isolated scattering events, that furthermore do not perturb the bath density operator. The impact of scattering events are modeled from the average distribution of scatterers, and ultimately lead to the transformation of the propagation operator,
\begin{align}
\mathrm{tr} \left\{ \hat{U}^{\dagger}(t) \hat{R}_{j\pm} \hat{U}(t)\ \hat{\rho}_0^{(M)} \otimes \hat{\rho}_0^{(B)}\right\} \\ \nonumber \to \mathrm{tr} \left\{ \hat{A}^{\dagger}(t) \hat{R}_{j\pm} \hat{A}(t)\ \hat{\rho}_0^{(M)} \right\},
\end{align}
to a non-unitary operator $\hat{A}(t)$ \cite{baranger:58, fano:63, fiutak:62, omont:72, bommier:97a}. We have put, as did \citet{fiutak:62, omont:72, bommier:97a}, the matrix-elements of the $\hat{A}(t)$ operators at, 
\begin{align}
\label{eq:eval_A}
\braket{a| \hat{A}^{\dagger}(t) | a'} \braket{b|\hat{A}(t)|b'} = \braket{ab||\hat{A}(t)||a'b'} = \delta_{aa'} \delta_{bb'} e^{-t/\tau_{ab}},
\end{align}
where elements of $\hat{A}(t)$ in Liouville space reduce to an exponential decorrelation of the characteristic time scale, $\tau_{ab}^{-1}$. Formally, the characteristic time scale is related to the transmission probability of the (averaged) scattering events, 
\begin{align}
\tau_{ab}^{-1} \propto 1 - \braket{a|\hat{S}|a} \braket{b|\hat{S}|b}^*,
\end{align}
where $\hat{S}$ is the scattering operator of the molecule with the perturber gas, averaged over all configurations. It is interesting to note here, that elastic collisions will only change the phase of the molecular states, while inelastic collisions will change the amplitude \cite{fiutak:62},
\begin{align*}
\braket{a|\hat{S}|a}_{\mathrm{el}} &= e^{i \eta_a}, \\
\braket{a|\hat{S}|a}_{\mathrm{inel}}  &= \sqrt{1-P_a},
\end{align*}
where $\eta_a$ is the (real) scattering phase and $P_a<1$ is a scattering probability \cite{fiutak:62}. Thus, $\tau_{ab}$ only contains a contribution from elastic scattering events only when $a \neq b$. Equation~(\ref{eq:eval_A}) allows us to derive the time-dependence of the operators,
\begin{subequations}
\label{eq:A_matrixelements}
\begin{align}
\hat{A}^{\dagger}(t) \hat{R}_{j\pm} \hat{A}(t) &= \hat{R}_{j\pm}\ e^{-t/\tau_{\pm \mp}}, \\
\hat{A}^{\dagger}(t) \hat{R}_{j\pm} \hat{R}_{j\mp} \hat{A}(t) &= \hat{R}_{j\pm} \hat{R}_{j\mp}\ e^{-t/\tau_{\pm \pm}}.
\end{align}
\end{subequations}
In our analysis, we have neglected any line shift due to collisional interactions, so that $\tau_{+-} = \tau_{-+}=\tau_L$. Furthermore, in the interest of simplicity, we assume $\tau_{++} = \tau_{--} = \tau_M$. In the literature on superradiance, $\tau_L$ and $\tau_M$ are better known as the characteristic dephasing time scale $T_2$ and the relaxation time scale $T_1$ \cite{rajabi:20}.

\subsection{Radiative transfer equation}
Now that we have established the interaction Hamiltonian, we turn to evaluate the transfer of radiation in the system under consideration. In order to get to a radiative transfer equation, we analyze the time-dependence of the radiation field specific intensity, that we track through the operator $\hat{i}_{\boldsymbol{k}} = [2h\nu/\lambda^2]\ \hat{a}_{\boldsymbol{k}}^{\dagger} \hat{a}_{\boldsymbol{k}}$. The operator, $\hat{i}_{\boldsymbol{k}}$, remains the same in the interaction picture, so we can note its time-dependence, 
\begin{align}
\label{eq:prop_in}
\frac{d}{dt} I_{\boldsymbol{k}}=
\frac{d}{dt} \mathrm{tr} \left\{\hat{i}_{\boldsymbol{k}} \hat{\rho}_I(t)\right\} = \mathrm{tr} \left\{\hat{i}_{\boldsymbol{k}} \frac{d\hat{\rho}_I(t)}{dt} \right\},
\end{align}
where we made use of the fact that $\frac{d\hat{i}_{\boldsymbol{k}}}{dt} = 0$. To proceed, we integrate Eq.~(\ref{eq:dens_interaction}) to yield
\begin{align}
\label{eq:densI_prop}
\frac{d\hat{\rho}_I(t)}{dt} &= -i \hbar^{-1} [\hat{V}(t),\hat{\rho}_0] \nonumber \\ &- \hbar^{-2} \int_0^t dt'\ [\hat{V}(t),[\hat{V}(t'),\hat{\rho}_I(t')]],
\end{align}
whilst making limiting, but physically reasonable assumptions, about the initial conditions of the transfer of radiation: 
\begin{subequations}
\label{eq:rad_ass}
\begin{align}
&\hat{\rho}_0 = \hat{\rho}_0^{(M)} \otimes \hat{\rho}_0^{(R)} \otimes \hat{\rho}_0^{(K)} \otimes \hat{\rho}_0^{(B)}, \\
&\mathrm{tr}\left\{ \hat{a}_{\boldsymbol{k}} \hat{\rho}_0 \right\}
= \mathrm{tr}\left\{ \hat{a}_{\boldsymbol{k}}^{\dagger}\hat{\rho}_0 \right\}
= 0 \\  &\mathrm{tr}\left\{ \hat{a}_{\boldsymbol{k}} \hat{a}_{\boldsymbol{k'}} \hat{\rho}_0 \right\} =
\mathrm{tr}\left\{ \hat{a}_{\boldsymbol{k}}^{\dagger} \hat{a}_{\boldsymbol{k'}}^{\dagger} \hat{\rho}_0 \right\} =0, \\
&\mathrm{tr}\left\{ \hat{a}_{\boldsymbol{k}'}^{\dagger} \hat{a}_{\boldsymbol{k}} \hat{\rho}_0 \right\} = 0, \qquad \mathrm{if} \ \boldsymbol{k} \neq \boldsymbol{k}'.
\end{align}
\end{subequations}
These assumptions amount to taking the radiation field, molecular ensemble, and bath as uncorrelated at $t=0$, as well as assuming that the incoming radiation field obeys Gaussian statistics.

We now combine the assumptions of Eqs.~(\ref{eq:rad_ass}), with Eqs.~(\ref{eq:prop_in}) and (\ref{eq:densI_prop}). First, we recognize that the element $[\hat{V}(t),\hat{\rho}_0]$, that appears in Eq.~(\ref{eq:densI_prop}) equals zero under the assumptions of Eqs.~(\ref{eq:rad_ass}). Using the usual commutation relations radiation field operators, we can parse out the commutator,
\[
\left[[\hat{i}_{\boldsymbol{k}},\hat{V}(t)],\hat{V}(t') \right] =\left[[\hat{i}_{\boldsymbol{k}},\hat{B}(t)],\hat{B}^{\dagger}(t') \right] + \left[[\hat{i}_{\boldsymbol{k}},\hat{B}^{\dagger}(t)],\hat{B}(t') \right],
\]
where under our line broadening assumptions, the second part is the complex conjugate of the first. Developing explicitly
\begin{align}
-&\hbar^{-2}\left[[\hat{i}_{\boldsymbol{k}},\hat{B}(t)],\hat{B}^{\dagger}(t') \right] \nonumber \\ &=  \sum_{jl\ \boldsymbol{k}'} g_{\boldsymbol{k}} g_{\boldsymbol{k}'}\frac{2 h\nu}{\lambda^2} e^{i(\omega_0-\omega_k)t}e^{-i(\omega_0-\omega_{k'})t'} e^{i \boldsymbol{k} \cdot \hat{\boldsymbol{v}}_j t} e^{-i \boldsymbol{k}' \cdot \hat{\boldsymbol{v}}_l t'}  \nonumber \\ &\times e^{i \boldsymbol{k} \cdot \hat{\boldsymbol{r}}_j} e^{-i \boldsymbol{k}' \cdot \hat{\boldsymbol{r}}_l}\left[ \tilde{R}_{j+}(t) \hat{a}_{\boldsymbol{k}},\tilde{R}_{l-}(t') \hat{a}_{\boldsymbol{k}'}^{\dagger} \right],
\end{align}
and using that the commutator of two products of operators, $\hat{A}\hat{X}$, and $\hat{B}\hat{Y}$, that commute as $[\hat{A},\hat{X}]=[\hat{B},\hat{Y}]=[\hat{A},\hat{Y}]=[\hat{B},\hat{X}]=0$, is \cite{landi:83}
\begin{align}
\left[\hat{A}\hat{X},\hat{B}\hat{Y}\right] = [\hat{A},\hat{B}] \hat{Y}\hat{X} + \hat{A}\hat{B} \left[\hat{X},\hat{Y}\right], \nonumber
\end{align}
we retrieve 
\begin{align}
\label{eq:rt_double_exp}
-&\hbar^{-2} \left[[\hat{i}_{\boldsymbol{k}},\hat{B}(t)],\hat{B}^{\dagger}(t') \right] = \sum_{j \boldsymbol{k}'} g_{\boldsymbol{k}} g_{\boldsymbol{k}'} [\tilde{R}_{j+}(t),\tilde{R}_{j-}(t')] \nonumber \\ &\times \left[\frac{2 h\nu}{\lambda^2} \hat{a}_{\boldsymbol{k}'}^{\dagger} \hat{a}_{\boldsymbol{k}} \right] \ e^{i(\omega_0-\omega_k)t} e^{-i(\omega_0-\omega_{k'})t'} e^{i \boldsymbol{k} \cdot \hat{\boldsymbol{v}}_j t} e^{-i \boldsymbol{k}' \cdot \hat{\boldsymbol{v}}_j t'}\nonumber \\ &\times e^{i (\boldsymbol{k}-\boldsymbol{k}') \cdot \hat{\boldsymbol{r}}_j} + \sum_{jl} \frac{2 h\nu}{\lambda^2} g_{\boldsymbol{k}}^2  \tilde{R}_{j+}(t) \tilde{R}_{l-}(t') \nonumber \\ &\times e^{i(\omega_0-\omega_k)(t-t')} e^{i \boldsymbol{k} \cdot \hat{\boldsymbol{v}}_jt} e^{-i \boldsymbol{k} \cdot \hat{\boldsymbol{v}}_lt'} e^{i \boldsymbol{k} \cdot (\hat{\boldsymbol{r}}_j - \hat{\boldsymbol{r}}_l)}.
\end{align}
In the interest of keeping the equations tractable, we will put the term $e^{i (\boldsymbol{k}-\boldsymbol{k}') \cdot \hat{\boldsymbol{r}}_j} \to \delta_{\boldsymbol{k},\boldsymbol{k}'}$. At first glance, such an operation appears an excellent approximation due to the ensemble being contained in a dimension much larger than the wavelength $L \gg \lambda$. Yet formally, it is too early to apply this operation, as correction terms due to the time-dependence of $\hat{\rho}_I(t)$ may affect the $e^{i (\boldsymbol{k}-\boldsymbol{k}')\cdot \hat{\boldsymbol{r}}_j}$ factor. Indeed, these do for the single-body corrections, and this will lead to single-body terms to be missing, when expanding the time-dependence. In contrast, for the many-body correction terms, the factor $e^{i (\boldsymbol{k}-\boldsymbol{k}')\cdot \hat{\boldsymbol{r}}_j}$ will be not be affected and remain invariant. Since we focus here on the many-body expansion terms (for a definition of the relevant parameter regime, see discussion later on), and only wish to know the order of magnitude of the single-body expansion terms, we will put $e^{i (\boldsymbol{k}-\boldsymbol{k}') \cdot \hat{\boldsymbol{r}}_j} \to \delta_{\boldsymbol{k},\boldsymbol{k}'}$ from here on. In Appendix~B, we verify the validity of this approximation.

Using the above results, adding the c.c.~to Eq.~(\ref{eq:rt_double_exp}), we can derive that the change in radiation specific intensity, per unit length, $ds = c dt$, is described according to the equation,
\begin{subequations}
\label{eq:lrt_dens}
\begin{align}
\frac{d I_{\boldsymbol{k}}}{ds} &= \int_0^{t} dt'\ \mathrm{tr} \left\{ \left[ \hat{\sigma}(t,t') + \hat{\gamma}(t,t') \right] \hat{\rho}_I (t') \right\}, \\
\hat{\sigma}(t,t') &= \frac{g_{\boldsymbol{k}}^2}{c} \sum_{j} [\tilde{R}_{j+} (t),\tilde{R}_{j-} (t')] \hat{i}_{\boldsymbol{k}} \nonumber \\  &\times  2\mathrm{Re}[e^{i(\omega_0-\omega_k)(t-t')} e^{i \boldsymbol{k}\cdot \hat{\boldsymbol{v}}_j(t-t')}], \\
\hat{\gamma} (t,t') &= \frac{2 h\nu g_{\boldsymbol{k}}^2}{\lambda^2 c}\sum_{j l} \tilde{R}_{j+}(t) \tilde{R}_{l-} (t')\nonumber \\  &\times 2\mathrm{Re}[e^{i \boldsymbol{k}\cdot (\boldsymbol{r}_j - \boldsymbol{r}_l)} e^{i(\omega_0-\omega_k)(t-t')} e^{i \boldsymbol{k}\cdot \hat{\boldsymbol{v}}_jt} e^{-i \boldsymbol{k}\cdot \hat{\boldsymbol{v}}_lt'}].
\end{align}
\end{subequations}
Equations~(\ref{eq:lrt_dens}) are exact but for three approximations. First, we assume that incoming radiation, at $t=0$, exhibits Gaussian statistics. Second, we use the rotating wave approximation. Third, it is assumed that dominant decorrelation of our system proceeds through the transition-level state operators, that are accordingly endowed with a time dependence. 

The formal solution to the radiative transfer equation of Eq.~(\ref{eq:lrt_dens}) may be found by using Eq.~(\ref{eq:densI_exact}), the cyclic property of the trace, and transforming the radiative operators, to leave a dependence on the initial density-operator only,
\begin{subequations}
\label{eq:rad_trans_ops}
\begin{align}
\label{eq:rad_trans_exact}
\frac{d I_{\boldsymbol{k}}}{ds} &= \int_0^t dt'\ \mathrm{tr} \left\{ \tilde{\mathcal{T}} e^{i \hbar^{-1} \int_0^{t'} dt'' \hat{V}(t'')}\left[ \hat{\sigma} (t,t') + \hat{\gamma}(t,t')\right] \right.  \nonumber \\ &\times \left. \hat{\mathcal{T}} e^{-i \hbar^{-1} \int_0^{t'} dt'' \hat{V}(t'')} \rho_0 \right\}.
\end{align}
Subsequently, the radiative operators can be expanded using \cite{fano:57},
\begin{align}
\label{eq:q_exp}
&\tilde{\mathcal{T}} e^{i \hbar^{-1} \int_0^{t'} dt'' \hat{V}(t'')} \hat{Q}(t,t') \hat{\mathcal{T}} e^{-i \hbar^{-1} \int_0^{t'} dt'' \hat{V}(t'')} = \nonumber \\ &\hat{Q}(t,t') - i\hbar^{-1} \int_0^{t'}dt'' \left[\hat{Q}(t,t'),\hat{V}(t'') \right] \nonumber \\ 
&- \hbar^2\int_0^{t'}dt'' \int_0^{t''}dt''' \left[\left[\hat{Q}(t,t'),\hat{V}(t'')\right],V(t''')\right] + \cdots ,
\end{align}
where we use $\hat{Q}$ as a stand-in for the radiative operators $\hat{\gamma}$ and $\hat{\sigma}$. Terms with an uneven number of $\hat{V}(t)$ operators fall away due to the conditions of Eq.~(\ref{eq:rad_ass}). In the remainder of the paper, we make use of shorthand notations
\begin{align}
&\tilde{\mathcal{T}} e^{i \hbar^{-1} \int_0^{t'} dt'' \hat{V}(t'')} \hat{\sigma}(t,t') \hat{\mathcal{T}} e^{-i \hbar^{-1} \int_0^{t'} dt'' \hat{V}(t'')} = \nonumber \\ &\hat{\sigma}_0(t,t') + \int_0^{t'}dt'' \int_0^{t''}dt''' \hat{\sigma}_1 (t,t',t'',t''') + \cdots \\
&\tilde{\mathcal{T}} e^{i \hbar^{-1} \int_0^{t'} dt'' \hat{V}(t'')} \hat{\gamma}(t,t') \hat{\mathcal{T}} e^{-i \hbar^{-1} \int_0^{t'} dt'' \hat{V}(t'')} = \nonumber \\ &\hat{\gamma}_0(t,t') + \int_0^{t'}dt'' \int_0^{t''}dt''' \hat{\gamma}_1 (t,t',t'',t''') + \cdots ,
\end{align}
and the formally traced time-integrals are noted
\begin{align}
\int_0^t& dt'\ \mathrm{tr} \left\{\tilde{\mathcal{T}} e^{i \hbar^{-1} \int_0^{t'} dt'' \hat{V}(t'')} \hat{\sigma}(t,t') \hat{\mathcal{T}}e^{-i \hbar^{-1} \int_0^{t'} dt'' \hat{V}(t'')}  \hat{\rho}_0\right\}  \nonumber \\ &= \sum_{n'} \sigma_{n'} \\
\int_0^t& dt'\ \mathrm{tr} \left\{\tilde{\mathcal{T}} e^{i \hbar^{-1} \int_0^{t'} dt'' \hat{V}(t'')} \hat{\gamma}(t,t') \hat{\mathcal{T}}e^{-i \hbar^{-1} \int_0^{t'} dt'' \hat{V}(t'')}  \hat{\rho}_0\right\} \nonumber \\ &= \sum_{n'} \gamma_{n'}.
\end{align}
\end{subequations}
\section{Many-body corrections to the radiative transfer equation}
Now, we set out to recover the canonical radiative transfer equation from the many-body formulation derived previously, and subsequently develop the corrections that arise when coherent interactions between molecules become non-negligible. In subsection A, we show how the standard form of the radiative transfer equation emerges under the assumption of weak coupling and short coherence times. In subsection B, we relax these assumptions, and evaluate the first-order corrections to the stimulated and spontaneous radiative operators, separating contributions from single-body and many-body interactions, and evaluating under what conditions the many-body terms dominate. In subsection C, we specialize to the regime where corrections from many-body interactions dominate, and extend the analysis to higher-order terms. We show that the many-body corrections form an exponential series, that ultimately lead to the modified radiative transfer equation of Eq.~(\ref{eq:lrt_mb}).
\subsection{Canonical radiative transfer equation}
In the previous section we derived the general radiative transfer equation of Eq.~(\ref{eq:lrt_dens}). We set out to investigate this equation in detail, but begin by making the simplest assumption, namely that the system evolves minimally under the interaction time, $\hat{\rho}_I (t) \approx \hat{\rho}_0$. This reduces Eq.~(\ref{eq:lrt_dens}) to
\begin{align*}
\left.\frac{dI_{\boldsymbol{k}}}{ds}\right|_0 &= \int_0^t dt' \ \mathrm{tr}\left\{ \left[\hat{\sigma}_0 (t,t') + \hat{\gamma}_0 (t,t')\right] \rho_0\right\} \\
&= \sigma_0 + \gamma_0,
\end{align*}
using the nomenclature introduced in Eqs.~(\ref{eq:rad_trans_ops}). First, we evaluate stimulated radiative operator,
\begin{align}
&\sigma_0 = \mathrm{tr} \left\{ \hat{\sigma}_0 (t,t') \hat{\rho}_0 \right\} = \frac{g_{\boldsymbol{k}}^2}{c} \sum_{j} \nonumber \\ &\mathrm{tr}\left\{ [\tilde{R}_{j+} (t),\tilde{R}_{j-} (t')]\hat{\rho}_0^{(M)} \otimes \hat{\rho}_0^{(B)} \right\} \mathrm{tr}\left\{\hat{i}_{\boldsymbol{k}} \hat{\rho}_0^{(R)} \right\}  \nonumber \\  &\times 2\mathrm{Re}\left[e^{i(\omega_0-\omega_k)(t-t')} \mathrm{tr}\left\{ e^{i \boldsymbol{k}\cdot \hat{\boldsymbol{v}}_j(t-t')}\hat{\rho}_0^{(K)} \right\}\right] \nonumber \\ &=
\frac{h \nu}{4\pi} B_0\frac{N}{V} (n_+ - n_-) I_{\boldsymbol{k}} \nonumber \\ &\times 2 \mathrm{Re}\left[ e^{i (\omega_0 - \omega_k)(t-t')} e^{-(t-t')/\tau_L} e^{- \left[\frac{t-t'}{\tau_D}\right]^2} \right], 
\end{align}
where we used $g_{\boldsymbol{k}}^2/c = h\nu B_0/ 4\pi V$, with $B_0$ as the Einstein $B$-coefficient and we have taken the particle velocities from a Maxwell-Boltzmann distribution as in Eq.~(\ref{eq:doppler_dec}). Furthermore, we used $\mathrm{tr} \{\hat{i}_{\boldsymbol{k}}\hat{\rho}_0^{(R)}\}=I_{\boldsymbol{k}}$, and we evaluated the molecular state operators, including their decorrelation, by using (Eq.~\ref{eq:collision_decor}, case $n=1$. See also, \citet{baranger:58, omont:72}),
\begin{align*}
\mathrm{tr} &\left\{\tilde{R}_{j+}(t_1) \tilde{R}_{l-} (t_2) \hat{\rho}_0^{(M)} \otimes \hat{\rho}_0^{(B)}\right\} \\ &= e^{-(t_1-t_2)/\tau_L} \mathrm{tr} \left\{\hat{R}_{j+} \hat{R}_{l-} \hat{\rho}_0^{(M)}\right\},\end{align*}
with
$$\mathrm{tr} \left\{\hat{R}_{j \pm} \hat{R}_{l \mp} \hat{\rho}_0^{(M)}\right\} = n_{\pm} \delta_{jl}.$$ 
We also let the summation $\sum_j \to N$, which is possible because when $\hat{\rho}_I(t) \approx \hat{\rho}_0$, no two different molecules are correlated during the interaction.

To obtain the final propagation element, we have to perform the time-integral
\begin{align}
&\int_0^t dt' \ \mathrm{tr} \left\{ \sigma_0 (t,t') \hat{\rho}_0 \right\}= \frac{h \nu}{4\pi} B_0\mathcal{N} (n_+ - n_-) I_{\boldsymbol{k}} \nonumber \\ &\int_0^t dt'\ 2\mathrm{Re} \left[ e^{i (\omega_0 - \omega_k)(t-t')} e^{-(t-t')/\tau_L} e^{- \left[\frac{t-t'}{\tau_D}\right]^2} \right], 
\end{align}
where we defined the particle density $\mathcal{N}=N/V$. The time-integral can be evaluated by making the substitution $\tau = t-t'$, and assuming a long interaction time $t \gg \tau_D,\tau_L$, so that 
\begin{align}
&\int_0^t dt'\ 2 \mathrm{Re}\left[ e^{i (\omega_0 - \omega_k)(t-t')} e^{-(t-t')/\tau_L} e^{- \left[\frac{t-t'}{\tau_D}\right]^2} \right] \nonumber \\ &\to \int_0^{\infty} d\tau\ 2 \mathrm{Re}\left[e^{i (\omega_0 - \omega_k)\tau} e^{-\tau/\tau_L} e^{- \left[\frac{\tau}{\tau_D}\right]^2} \right] =\phi_{\nu},
\end{align}
yields a line profile that is centered around $\nu_0 = \omega_0/2\pi$. This is a Gaussian profile when decorrelation due to Doppler-broadening dominates, $\tau_D \ll \tau_L$, a Lorentzian when decorrelation due to lifetime and/or pressure broadening dominates, $\tau_L \ll \tau_D$, and a Voigt profile when both broadening mechanisms are of comparable importance. From now on, we focus on Doppler broadened lines. 

The spontaneous emission term is worked out similarly
\begin{align}
\gamma_0 =\int_0^t dt' \ \mathrm{tr} \left\{ \hat{\gamma}_0 (t,t') \hat{\rho}_0 \right\}= \frac{h \nu}{4\pi} A_0 \mathcal{N} n_+ \phi_{\nu} ,
\end{align}
where we introduced $A_0$ as the Einstein $A$-coefficient, that us related to the Einstein $B$-coefficient by the relation $[2 h \nu/\lambda^2]B_0 = A_0$.

Adding the stimulated and spontaneous radiative interaction terms, $\sigma_0$ and $\gamma_0$, yields the radiative transfer equation,
\begin{align}
\label{eq:rad_trans_0}
\left. \frac{d I_{\boldsymbol{k}}}{ds} \right|_0 &= -\frac{h \nu}{4\pi} B_0 \mathcal{N} (n_- - n_+) \phi_{\nu} I_{\boldsymbol{k}} + \frac{h \nu}{4\pi} A_0 \mathcal{N} n_+ \phi_{\nu} \nonumber \\ 
&=- \kappa_{\nu} I_{\boldsymbol{k}} + \epsilon_{\nu} .
\end{align}
This is the well-known and often used form of the line radiative transfer equation. Together with the statistical equilibrium equations for the level populations, the radiative transfer equation forms the cornerstone of nearly all line radiative transfer formalisms. At wavelengths longer than infrared, it is generally assumed that radiative transfer can be comprehensively described this way, while at shorter wavelengths often a correction due to line scattering is added.

\subsection{Corrections to the canonical radiative transfer equation}
In the previous subsection, we recovered the canonical radiative transfer equation in the limit of negligible system evolution under the interaction time. We now proceed to evaluate the leading-order corrections that arise when this approximation breaks down. We distinguish between single-body corrections, which involve only interactions of a single molecule with the radiation field, and many-body contributions, which capture collective effects due to the coherent interaction of multiple molecules. We compare their magnitudes, in order to define the regimes where the many-body terms dominate. We show that in Doppler broadened media, many-body corrections to the stimulated radiative operator dominate over many-body corrections to the spontaneous radiative operator. 

\subsubsection{First-order correction to the stimulated radiative operator}
We begin by evaluating the leading-order corrections to the stimulated radiative operator. We recall Eqs.~(\ref{eq:rad_trans_ops}), and note that the leading-order corrections to the stimulated radiative operator correspond to
\begin{align}
\label{eq:kv_1_comm}
\hat{\sigma}_1 (t,t',t'',t''') &= -\hbar^{-2} [[\hat{\sigma}(t,t'),\hat{V}(t'')],\hat{V}(t''')],\end{align}
in its operator form, while $\sigma_1$ represents the time-integrated and traced element that is the first-order correction to the radiative transfer equation. First, we evaluate Eq.~(\ref{eq:kv_1_comm}), and afterwards, we perform its proper time-integration. We notice that the stimulated radiative operator, $\hat{\sigma}(t,t')$, contains the operator product, $2\hat{R}_{j3}(t,t') \hat{i}_{\boldsymbol{k}}$, where we recall that $2\hat{R}_{j3}(t,t') = [\hat{R}_{j+}(t),\hat{R}_{j-}(t')]$. Thus, we can parse out 
\begin{widetext}
\begin{align}
\label{eq:kv_1_comm_parsed}
&\hat{\sigma}_1 (t,t',t'',t''') =\left[g_{\boldsymbol{k}}^2 \hat{i}_{\boldsymbol{k}} \right] \left(-c^{-1}\hbar^{-2} \sum_j \left[[2\hat{R}_{{j3}} (t,t'),\hat{V}(t'')],\hat{V}(t''')\right] f_{j}(t-t')  \right) 
+\left[\sum_{j} 2 g_{\boldsymbol{k}}^2 \hat{R}_{j3}(t,t')  f_{j}^{\boldsymbol{k}}(t-t') \right] \nonumber \\ &\times\left(-c^{-1}\hbar^{-2} \left[[\hat{i}_{\boldsymbol{k}},\hat{V}(t'')],\hat{V}(t''')\right] \right) + \sum_{j} \left([-c^{-1} \hbar^{-2}g_{\boldsymbol{k}}^2 f_j^{\boldsymbol{k}}(t-t')] \left\{[\hat{i}_{\boldsymbol{k}},\hat{V}(t'')] [2\hat{R}_{j3},\hat{V}(t''')]  + [\hat{i}_{\boldsymbol{k}},\hat{V}(t''')] [2\hat{R}_{j3},\hat{V}(t'')]\right\} \right) \nonumber \\ &= \hat{\sigma}_1^{\mathrm{s.b.}} (t,t',t'',t''') + \hat{\sigma}_1^{\mathrm{m.b.}} (t,t',t'',t''') + \hat{\sigma}_1^{\mathrm{s.b.\ forward}} (t,t',t'',t'''),
\end{align}
\end{widetext}
where we use the shorthand notation $$f_{j}^{\boldsymbol{k}} (t-t') = 2\mathrm{Re}[ e^{i(\omega_0-\omega_k)(t-t')} e^{i \boldsymbol{k}\cdot \hat{\boldsymbol{v}}_j(t-t')}].$$
Note that we have called each the elements of the sum of Eq.~(\ref{eq:kv_1_comm_parsed}), $\hat{\sigma}_1^{\mathrm{s.b.}} (t,t',t'',t''') $, referring to the `single-body' corrections, $\hat{\sigma}_1^{\mathrm{m.b.}} (t,t',t'',t''') $, referring to the `many-body' corrections, and $\hat{\sigma}_1^{\mathrm{s.b.\ forward}} (t,t',t'',t''')$ referring to `single-body interactions in the forward direction'. We elect this nomenclature, because, when evaluated, the operator $\hat{\sigma}_1^{\mathrm{s.b.}}$ only returns non-zero values for the $j$'th particle contributing to the $\hat{V}(t'')$ operator, while it does allow other photon modes, $\boldsymbol{k}'$ to contribute to the interaction. In contrast, for the operator $\hat{\sigma}_1^{\mathrm{m.b.}} $, the forward direction, $\boldsymbol{k}=\boldsymbol{k}'$, is forced through the first commutator, but all particles in $\hat{V}(t'')$ contribute to the interaction. In other words, in the single-body interaction, the number of participating photon modes is maximized, while in the many-body interaction, the number of participating molecules is maximized. Single-body interactions in the forward direction have the limiting features of both the single-body and many-body interactions, and are therefore of low magnitude. 

We work out the single-body and many-body corrections individually and compare their magnitudes. Using the notation, $\tilde{R}_{j+}([t,t'];t'') = [[\tilde{R}_{j+}(t),\tilde{R}_{j-}(t')],\tilde{R}_{j+}(t'')]$, we proceed and work out the full commutator of the single-body interaction
\begin{widetext}
\begin{align}
\hat{\sigma}_1^{\mathrm{s.b.}} (t,t',t'',t''') &= \left[g_{\boldsymbol{k}}^2 \hat{i}_{\boldsymbol{k}} \right] \sum_{j l \boldsymbol{k}' \boldsymbol{k}''}\frac{g_{\boldsymbol{k}'} g_{\boldsymbol{k}''}}{c} [\tilde{R}_{j+}([t,t'];t'') \hat{a}_{\boldsymbol{k}'}, \tilde{R}_{l-}(t''')\hat{a}_{\boldsymbol{k}''}^{\dagger}]\nonumber \\ &\times f_j^{\boldsymbol{k}} (t-t') e^{i(\omega_0-\omega_{k'})t''} e^{-i(\omega_0-\omega_{k''})t'''} e^{i \boldsymbol{k}'\cdot \hat{\boldsymbol{r}}_{j}}e^{i \boldsymbol{k}'\cdot \hat{\boldsymbol{v}}_j t''} e^{-i \boldsymbol{k}''\cdot \hat{\boldsymbol{r}}_{l}} e^{-i \boldsymbol{k}''\cdot \hat{\boldsymbol{v}}_l t'''} + \mathrm{c.c.} \nonumber \\ 
&= \left[g_{\boldsymbol{k}}^2 \hat{i}_{\boldsymbol{k}} \right]  \sum_{j l \boldsymbol{k}' \boldsymbol{k}''} \frac{g_{\boldsymbol{k}'} g_{\boldsymbol{k}''}}{c}  \left([\tilde{R}_{j+}([t,t'];t'') , \tilde{R}_{l-}(t''')]  \hat{a}_{\boldsymbol{k}''}^{\dagger} \hat{a}_{\boldsymbol{k}'} + \tilde{R}_{j+}([t,t'];t'')  \tilde{R}_{l-}(t''') [\hat{a}_{\boldsymbol{k}'},\hat{a}_{\boldsymbol{k}''}^{\dagger}] \right)  \nonumber \\
&\times f_j^{\boldsymbol{k}} (t-t') e^{i(\omega_0-\omega_{k'})t''} e^{-i(\omega_0-\omega_{k''})t'''} e^{i \boldsymbol{k}'\cdot \hat{\boldsymbol{r}}_{j}} e^{i \boldsymbol{k}'\cdot \hat{\boldsymbol{v}}_j t''} e^{-i \boldsymbol{k}''\cdot \hat{\boldsymbol{r}}_{l}} e^{-i \boldsymbol{k}''\cdot \hat{\boldsymbol{v}}_l t'''}+ \mathrm{c.c.} . 
\end{align}
\end{widetext}
The first term in the sum can be simplified by recognizing that the commutator is only non-zero when $j = l$. While for the second term in the sum, we take an advance on the expectation value, $\mathrm{tr}\left\{ \hat{R}_{j+} \hat{R}_{l-} \hat{\rho}_0^{(M)} \right\} = \delta_{jl} n_+$, that also forces $j=l$. Then, we recognize furthermore that the phase-factor $e^{i(\boldsymbol{k}'-\boldsymbol{k}'')\cdot \hat{\boldsymbol{r}}_{j}} \to \delta_{\boldsymbol{k}',\boldsymbol{k}''}$ for the large sample that we consider. Together, this yields
\begin{align}
&\hat{\sigma}_1^{\mathrm{s.b.}} (t,t',t'',t''') = \frac{g_{\boldsymbol{k}}^2 \hat{i}_{\boldsymbol{k}}}{c} \sum_{j\boldsymbol{k}'} g_{\boldsymbol{k}'}^2 \left( [\tilde{R}_{j+}([t,t'];t''),\tilde{R}_{j-}(t''')] \right. \nonumber \\  &\times \hat{n}_{\boldsymbol{k}'} + \left. \tilde{R}_{j+}([t,t'];t'') \tilde{R}_{j-}(t''') \right) f_{j}^{\boldsymbol{k}} (t-t') f_{j}^{\boldsymbol{k}'} (t''-t'''),
\end{align}
which will be the form we will perform the time integrals on later on.

Now we proceed to consider the many-body interaction term. We derived the radiative transfer equation in Sec.~II.B.,~where we showed in detail that
$$
-c^{-1}\hbar^{-2} [[\hat{i}_{\boldsymbol{k}},\hat{V}(t'')],\hat{V}(t''')] = \hat{\sigma} (t'',t''') + \hat{\gamma} (t'',t'''),
$$
so 
\begin{align}
\hat{\sigma}_1^{\mathrm{m.b.}} (t,t',t'',t''') &= \left[\sum_j 2g_{\boldsymbol{k}}^2 \tilde{R}_{j3}(t,t') f_{j}^{\boldsymbol{k}}(t-t') \right] \nonumber \\ &\times \left[\hat{\sigma}(t'',t''') + \hat{\gamma}(t'',t''') \right].
\end{align}
Thus, the many-body correction to the stimulated radiative operator contains again the stimulated radiative operator.  Later on, we will show that for a Doppler broadened line, many-body correction to the stimulated radiative operator dominates over corrections to the spontaneous emission operator, so that we are allowed to use this relation as a recursion relation, and expand the many-body corrections to the stimulated radiative operator to arbitrary order.

Proceeding to evaluate the time-integrals of $\hat{\sigma}_1^{\mathrm{s.b.}} (t,t',t'',t''') $ and $\hat{\sigma}_1^{\mathrm{m.b.}} (t,t',t'',t''') $, we use results from Appendix~A, to assign time-decorrelation functions to the molecular operators. We make the substitutions $\tau = t-t'$, $\tau ' = t'-t''$ and $\tau'' = t''-t'''$. Molecular operators experience a decorrelation through interactions with the bath. This decorrelation during times $\tau$ and $\tau''$, is of characteristic time scale $\tau_L$, as during these times, a transition is ongoing, while during times $\tau'$, the system spends its time in either of the molecular states, so the characteristic decorrelation time scale is $\tau_M$. For more discussion on this, see \citet{fiutak:62} and Appendix~A.

With these additional decorrelation functions, and under the assumption $t \gg \tau_L, \tau_M, \tau_D$, the time integrals can be evaluated. This allows us to compute the first-order corrections to the stimulated radiative operator in the radiative transfer equation. We carry out the evaluation separately for the single-body and many-body contributions, beginning with the single-body case.
\begin{widetext}
\begin{align}
\label{eq:kv_nc_2}
&\int_0^t dt'\int_0^{t'}dt'' \int_0^{t''}dt'''\ \mathrm{tr}\left\{ \hat{\sigma}_1^{\mathrm{s.b.}} (\boldsymbol{k};t,t',t'',t''') \hat{\rho}_0\right\} 
\nonumber \\
&\simeq -\left(\frac{h\nu}{4\pi} B_0 \mathcal{N} (n_+ - n_-) \right) I_{\boldsymbol{k}} \sum_{\boldsymbol{k'}}\left[ g_{\boldsymbol{k}'}^2 [2h\nu/\lambda^2]^{-1}I_{\boldsymbol{k}'} + \frac{n_+}{n_+ - n_-} g_{\boldsymbol{k}'}^2  \right] \nonumber \\ &\times \int_0^{\infty} d\tau\ 2\mathrm{Re}\left[ e^{i(\omega_0 - \omega_k) \tau} e^{-\tau^2 / \tau_D^2} e^{-\tau / \tau_L} \right]\int_0^{\infty} d\tau'' 2\mathrm{Re}\left[e^{i(\omega_0 - \omega_k') \tau''} e^{-\tau''^2 / \tau_D^2} e^{-\tau'' / \tau_L}\right]\int_0^{\infty}  d \tau'\  e^{-\tau' / \tau_M}\nonumber \\ 
&= -\left(\frac{h\nu}{4\pi} B_0 \mathcal{N} (n_+ - n_-) \phi_{\nu}\right) I_{\boldsymbol{k}} \sum_{\boldsymbol{k'}}\left[ g_{\boldsymbol{k}'}^2 [2h\nu/\lambda^2]^{-1}I_{\boldsymbol{k}'} \phi_{\nu'} + \frac{n_+}{n_+ - n_-} g_{\boldsymbol{k}'}^2 \phi_{\nu'} \right]\tau_{\mathrm{M}} \nonumber \\ &= 
-\kappa_{\nu} I_{\boldsymbol{k}} \left[-\frac{B_0 \bar{i}}{2 \tau_{\mathrm{M}}^{-1}} + \frac{A_0}{2 \tau_{\mathrm{M}}^{-1}} \frac{n_+}{n_- - n_+} \right],
\end{align}
\end{widetext}
where we substituted the summation over all photon modes by
\[
\sum_{\boldsymbol{k}} \to \frac{V}{c^3} \int d\nu_{k}\ \nu_{k}^2 \int d\hat{k},
\]
when defining the line profile integrated intensity,
\[
\bar{i} = \int d\nu_k \phi_{\nu_k} \int \frac{d\hat{k}}{4\pi}\ I_{\boldsymbol{k}},
\]
where we furthermore used that the integrand over the frequency is sharply peaked around $\nu_0$, so that the factor $\nu_k^2 \approx \nu_0^2$, and can be taken out of the integrand. We have averaged the particle velocities to yield the Doppler decorrelation as in Eq.~(\ref{eq:doppler_dec}). 
We call $B_0 \bar{i}=A_0 \bar{n}$, the rate of stimulated emission and absorption.

Now, we proceed to analyze the many-body contribution,
\begin{widetext}
\begin{align}
\label{eq:kv_c_2}
&\int_0^t dt'\int_0^{t'}dt'' \int_0^{t''}dt'''\ \mathrm{tr}\left\{ \hat{\sigma}_1^{\mathrm{m.b.}} (t,t',t'',t''') \hat{\rho}_0\right\} \nonumber \\
&\simeq \left(\frac{h\nu}{4\pi} B_0 \mathcal{N} (n_+ - n_-) c \right) \left[\frac{h \nu}{4\pi} B_0 \mathcal{N}(n_+ - n_-) I_{\boldsymbol{k}} + \frac{h\nu}{4\pi} A_0 \mathcal{N} n_+ \right]  \nonumber \\ 
&\times \int_0^{\infty} d\tau\ 2\mathrm{Re}\left[ e^{i(\omega_0 - \omega_k) \tau} e^{-[\tau / \tau_D]^2} e^{-\tau / \tau_L} \right]\int_0^{\infty} d\tau''\ 2\mathrm{Re}\left[e^{i(\omega_0 - \omega_k) \tau''} e^{-[\tau'' / \tau_D]^2} e^{-\tau'' / \tau_L}\right]\int_0^{\infty}  d \tau'\  e^{-\tau' / \tau_M} \nonumber \\ 
&= \left(\frac{h\nu}{4\pi} B_0 \mathcal{N} (n_+ - n_-) [c \tau_{\mathrm{M}}] \phi_{\nu}\right) \left[\frac{h \nu}{4\pi} B_0 \mathcal{N}(n_+ - n_-) I_{\boldsymbol{k}}\phi_{\nu} + \frac{h\nu}{4\pi} A_0 \mathcal{N} n_+ \phi_{\nu}\right]  \nonumber \\ 
&= -\tau_{\nu}^{\mathrm{coh}} \left[\epsilon_{\nu} - \kappa_{\nu} I_{\boldsymbol{k}} \right],
\end{align}
\end{widetext}
that returns the canonical radiative transfer equation, times the coherent optical depth. 

To evaluate the relative magnitude of the single-body and many-body corrections to the stimulated radiative operators, we compare Eqs.~(\ref{eq:kv_nc_2}) and (\ref{eq:kv_c_2}). Using that $\phi_{\nu}\sim \tau_D$, we obtain, $$\frac{\tau_{\nu}^{\mathrm{coh}}}{A_0(\bar{n}+1)\tau_M/2}\sim \frac{\lambda^2 \ell_D \mathcal{N}}{\bar{n}+1},$$
where $\ell_D = c \tau_D$, as the relative magnitude of many-body and single-body second-order interactions. 

We remark that if the rate of stimulated emission and absorption is large, relative to the state lifetimes, $B_0 \bar{i} / 2 \tau_M^{-1}>1$, then extending the single-body corrections to infinite order leads to a divergent series. In such regimes, the transfer of radiation can no longer be accurately described by Eq.~(\ref{eq:lrt_can}) or Eq.~(\ref{eq:lrt_mb}), because the assumption of steady-state, implicit in the time-integrals extending to infinite time, is violated. Instead, a more appropriate formalism is based on the Maxwell–Bloch equations \cite{menegozzi:78, sargent:74, lehmberg:70}. This distinction explains why many previous investigations of many-body radiative transfer have emphasized high-intensity systems, where such non steady-state approaches are necessary \cite{lehmberg:70, arecchi:70, gross:82}. In contrast, our work targets the low-intensity regime, $B_0 \bar{i} / 2 \tau_M^{-1}<1$, where, as will become apparent, many-body corrections remain tractable, and can be quantified in a steady-state formalism.

\subsubsection{First-order many-body corrections: spontaneous radiative operator vs stimulated radiative operator}
Having addressed the stimulated radiative operator, we next consider the spontaneous radiative operator. We evaluate the first-order many-body correction to assess the relative importance of many-body corrections from spontaneous and stimulated radiative interactions. We recall Eqs.~(\ref{eq:rad_trans_ops}), and start by working out the spontaneous radiative operator term,
\begin{widetext}\begin{align}
\label{eq:gamma_1_op}
\hat{\gamma}_1 (t,t',t'',t''') &= -\hbar^{-2} \left[[\hat{\gamma}(\boldsymbol{k};t,t'),\hat{V}(t'')],\hat{V}(t''')\right] \nonumber \\
&=-\left[\frac{2h\nu}{\lambda^2} \frac{g_{\boldsymbol{k}}^2}{c} \right] \sum_{jlmn}\sum_{\boldsymbol{k}'} g_{\boldsymbol{k}'}^2 \left\{\left[[\tilde{R}_{j+}(t) \tilde{R}_{l-}(t'),\tilde{R}_{m+} (t'') a_{\boldsymbol{k'}}], \tilde{R}_{n-}(t''') a_{\boldsymbol{k'}}^{\dagger}\right] h_{j} (\boldsymbol{k};t) h_{l}^*(\boldsymbol{k};t') h_{m}(\boldsymbol{k}';t'') h_{n}^*(\boldsymbol{k}';t''') \right.\nonumber \\
&+ \left. \left[[\tilde{R}_{j+}(t) \tilde{R}_{l-}(t'),\tilde{R}_{m-}(t'') a_{\boldsymbol{k'}}^{\dagger}],\tilde{R}_{n+} (t''') a_{\boldsymbol{k'}}\right] h_{j}^* (\boldsymbol{k};t) h_{l}(\boldsymbol{k};t') h_{m}^*(\boldsymbol{k}';t'') h_{n}(\boldsymbol{k}';t''')\right\} ,
\end{align}\end{widetext}
where we forced $\boldsymbol{k}'=\boldsymbol{k}''$, in anticipation of the trace in $\gamma_1$, and we gave a shorthand notation to the time-functions
\[
h_j (\boldsymbol{k};t) = e^{i (\omega_0 - \omega_k)t} e^{i \boldsymbol{k}\cdot \hat{\boldsymbol{v}}_jt}  e^{i \boldsymbol{k}\cdot \hat{\boldsymbol{r}}_j }.
\]
Both commutators can be worked out to yield the similar operators
\begin{align}
&\left[[\tilde{R}_{j+}(t) \tilde{R}_{l-}(t'),\tilde{R}_{m+}(t'') a_{\boldsymbol{k'}} ], \tilde{R}_{n-}(t''') a_{\boldsymbol{k'}}^{\dagger}\right] \nonumber \\ &=\delta_{lm} [\tilde{R}_{l-}(t'),\tilde{R}_{l+}(t'')] [\tilde{R}_{j+}(t) a_{\boldsymbol{k}'},\tilde{R}_{n-} (t''')a_{\boldsymbol{k}'}^{\dagger}] \nonumber \\
&\left[[\tilde{R}_{j+}(t) \tilde{R}_{l-}(t'),\tilde{R}_{m-}(t'') a_{\boldsymbol{k'}}^{\dagger} ], \tilde{R}_{n+} (t''') a_{\boldsymbol{k'}}\right] \nonumber \\ &= \delta_{jm} [\tilde{R}_{j+}(t),\tilde{R}_{j-}(t'')] [\tilde{R}_{l-} (t') a_{\boldsymbol{k}'}^{\dagger},\tilde{R}_{n+} (t''') a_{\boldsymbol{k}'}] \nonumber,
\end{align}
where the latter commutator we were already familiar with, and reduces to
\[
[\hat{R}_{j+}a_{\boldsymbol{k}'},\hat{R}_{n-}a_{\boldsymbol{k}'}^{\dagger}] = \hat{R}_{j+}\hat{R}_{n-} + \delta_{jn}[\hat{R}_{j+},\hat{R}_{j-}] \hat{n}_{\boldsymbol{k}'}.
\]
Having evaluated these commutators, we anticipate the eventual time functions that need to be integrated. We focus on the product, $h_{j} (\boldsymbol{k};t) h_{l}^*(\boldsymbol{k};t') h_{m}(\boldsymbol{k}';t'') h_{n}^*(\boldsymbol{k}';t''')$, in the first term in the sum of Eq.~(\ref{eq:gamma_1_op}). We use the above results to put $j=n$ and $l=m$ for this term, and work out explicitly the total time-dependent function to be
\begin{align*}
&h_{j} (\boldsymbol{k};t) h_{l}^*(\boldsymbol{k};t') h_{l}(\boldsymbol{k}';t'') h_{j}^*(\boldsymbol{k}';t''') = e^{i(\omega_0-\omega_k)(t-t')} \nonumber \\ &\times e^{i(\omega_0-\omega_{k'})(t''-t''')} e^{i(\boldsymbol{k}t - \boldsymbol{k}'t''')\cdot \hat{\boldsymbol{v}}_j}  e^{-i(\boldsymbol{k}t' - \boldsymbol{k}'t'')\cdot \hat{\boldsymbol{v}}_l} \nonumber \\ &\times e^{i(\boldsymbol{k}-\boldsymbol{k}')\cdot (\hat{\boldsymbol{r}}_j - \hat{\boldsymbol{r}}_l)}      
\end{align*}
Eventually, the final term $e^{i(\boldsymbol{k}-\boldsymbol{k}')\cdot (\hat{\boldsymbol{r}}_j - \hat{\boldsymbol{r}}_l)}$ will be evaluated, which yields for our sample of large extent, either the forcing of the forward direction, $\delta_{\boldsymbol{k,k'}}$, which is the (many-body) term that maximizes the amount of participating molecules in the interaction, or the forcing of the same molecule, $\delta_{jl}$, which is the (single-body) term that maximizes the participating photon modes in the interaction.

We focus here on the regime where many-body terms dominate over single-body terms, so we only consider the many-body term in the function
\begin{align*}
&h_{j} (\boldsymbol{k};t) h_{l}^*(\boldsymbol{k};t') h_{l}(\boldsymbol{k}';t'') h_{j}^*(\boldsymbol{k}';t''') \to e^{i(\omega_0-\omega_k)(t-t')} \nonumber \\ &\times e^{i(\omega_0-\omega_{k})(t''-t''')} e^{i\boldsymbol{k}\cdot \hat{\boldsymbol{v}}_j(t-t''')}  e^{-i\boldsymbol{k}\cdot \hat{\boldsymbol{v}}_l(t'-t'')} \delta_{\boldsymbol{k},\boldsymbol{k}'}   
\end{align*}
With the above results, we are ready to compute the first order many-body correction to the spontaneous radiative operator, $\gamma_1$. We do so, by first anticipating the time-integral that emerges in
\begin{align}
\label{eq:gamma_one_exp}
&\int_0^t dt'\int_0^{t'}dt'' \int_0^{t''}dt''' \mathrm{tr}\left\{\hat{\gamma}_1 (t,t',t'',t''') \hat{\rho}_0\right\}.\end{align}
We first compute the expectation value of the time-dependent function, that yields
\begin{align*}
&\mathrm{tr}\left\{h_{j} (\boldsymbol{k};t) h_{l}^*(\boldsymbol{k};t') h_{l}(\boldsymbol{k}';t'') h_{j}^*(\boldsymbol{k}';t''') \hat{\rho}_{0}^{(K)}\right\}  \nonumber \\ &= e^{i(\omega_0-\omega_k)(t-t')}e^{i(\omega_0-\omega_{k})(t''-t''')} \nonumber \\ &\times e^{-(t-t''')^2/\tau_D^2}  e^{-(t'-t'')^2/\tau_D^2} \delta_{\boldsymbol{k},\boldsymbol{k}'},
\end{align*}
where we used Eq.~(\ref{eq:doppler_dec}), and then proceed to evaluate the complete time-dependency integral that emerges in Eq.~(\ref{eq:gamma_one_exp})
\begin{widetext}
\begin{align}
&\int_0^t dt'\int_0^{t'}dt'' \int_0^{t''}dt'''\ 2\mathrm{Re}\left[ e^{i(\omega_0 - \omega_k)(t-t')} e^{i(\omega_0-\omega_k)(t''-t''')} e^{-[(t-t''')/\tau_D]^2} e^{-[(t'-t'')/\tau_D]^2} \right] \nonumber \\ &\times\mathrm{tr} \left\{\tilde{R}_{j\pm}(t) \tilde{R}_{j\pm'}(t') \tilde{R}_{j\pm''}(t'') \tilde{R}_{j\pm'''}(t''') \hat{\rho}_0^{(M)} \otimes \hat{\rho}_0^{(B)}\right\},
\end{align}
where we used the result that the second term in Eq.~(\ref{eq:gamma_1_op}) is the c.c.~of the first. Interestingly, we note that when substituting the in the usual way $\tau = t - t'$, $\tau' = t' - t''$ and $\tau'' = t'' - t'''$, then due to the time-dependencies, $e^{-[(t-t''')/\tau_D]^2} e^{-[(t'-t'')/\tau_D]^2}$, each time variable, $\tau$, $\tau'$ and $\tau''$, is associated with a decorrelation on the order of the characteristic Doppler time scale. Since we consider a system that is Doppler broadened, $\tau_D \ll \tau_L, \tau_M$, we note that the decorrelation of the particle operators under the interaction times should be negligible, so we proceed with the analysis assuming these operators are unaffected by decorrelation. We can then work out  
\begin{align}
&\int_0^t dt'\int_0^{t'}dt'' \int_0^{t''}dt'''\ \mathrm{tr}\left\{\hat{\gamma}_1 (t,t',t'',t''') \hat{\rho}_0\right\}
\nonumber \\
&= -\left[\frac{2h\nu}{\lambda^2} \frac{g_{\boldsymbol{k}}^4 (n_- - n_+) N^2}{c} \right] \left[ n_+ + (n_+-n_-) n_{\boldsymbol{k}} \right] \nonumber \\ &\times
\int_0^t dt'\int_0^{t'}dt'' \int_0^{t''}dt'''\ 2\mathrm{Re}\left[ e^{i(\omega_0 - \omega_k)(t-t')} e^{i(\omega_0-\omega_k)(t''-t''')} e^{-[(t-t''')/\tau_D]^2} e^{-[(t'-t'')/\tau_D]^2} \right].
\end{align}
We evaluate the time-integrals separately,
\begin{align}
 &\int_0^t dt'\int_0^{t'}dt'' \int_0^{t''}dt'''\ 2\mathrm{Re}\left[ e^{i(\omega_0 - \omega_k)(t-t')} e^{i(\omega_0-\omega_k)(t''-t''')} e^{-[(t-t''')/\tau_D]^2} e^{-[(t'-t'')/\tau_D]^2} \right] \nonumber \\   &\simeq \int_0^{\infty}d\tau \int_0^{\infty}d\tau' \int_0^{\infty}d\tau'' 2\mathrm{Re}\left[ e^{i(\omega_0 - \omega_k)\tau} e^{i(\omega_0-\omega_k)\tau''} e^{-[(\tau+\tau'+\tau'')/\tau_D]^2} e^{-[\tau'/\tau_D]^2} \right]\nonumber \\
&= \frac{\tau_D^3\sqrt{\pi}}{2\sqrt{2}} \int_{0}^{\infty} du \ u\, e^{-u^2 /2}  \cos \left[ (\omega_0-\omega_k)\tau_D\,u\right] \operatorname{erfc}\Bigl(\frac{u}{\sqrt{2}}\Bigr)  \nonumber \\
&= \phi_{\nu}^2 \frac{\tau_D}{2} \Psi_{\nu}.
\end{align}
\end{widetext}

While the previous integral is not solvable analytically, we have permitted ourselves to factorize it in regular line profiles $\phi_{\nu}$, an extra time-factor $\tau_D$ and a dimensionless profile $\Psi_{\nu}$ of $\mathcal{O}(1)$. In this way, we can return the second-order correction to the coherent spontaneous emission operator in a similar way as we did for the stimulated radiative operator, as
\begin{align}
\label{eq:tau_D_coh}
&\int_0^t dt'\int_0^{t'}dt'' \int_0^{t''}dt'''\ \mathrm{tr}\left\{\hat{\gamma}_1 (t,t',t'',t''') \hat{\rho}_0\right\} \nonumber \\
&= -\left[\frac{h\nu}{4\pi}  B_0 \mathcal{N} (n_- - n_+) \phi_{\nu} \Psi_{\nu} \frac{c \tau_D}{2} \right] \nonumber \\ &\times\left[ \frac{h\nu}{4\pi} A_0 \mathcal{N} n_+ \phi_{\nu} + \frac{h\nu}{4\pi} B_0 \mathcal{N} (n_+-n_-) \phi_{\nu}I_{\boldsymbol{k}} \right] \nonumber \\
&= -\tau_{\nu}^{D,\mathrm{coh}} \left[ \epsilon_{\nu} - \kappa_{\nu} I_{\boldsymbol{k}} \right].
\end{align}
Thus, in the first-order many-body correction to the spontaneous radiative operator, we have an equation similar to the full radiative transfer equation, but with a corrective multiplicative factor of $\tau_{\nu}^{D,\mathrm{coh}}$, which is a measure for the optical depth over a Doppler coherence length $c \tau_D$. 

Importantly, from this analysis we learned that many-body corrections to the spontaneous radiative operators, and stimulated radiative operator differ in the relevant decorrelation processes. While for the many-body stimulated radiative operator, these lie in the `survival' of molecules throughout the interaction process, for many-body spontaneous radiative operator, coherence has to be maintained between different molecules. This type of decorrelation is dominated by Doppler decorrelation, where the relative motion of two particles causes them to become decorrelated. Thus, the magnitude between both interactions scales as $\tau_M/\tau_D$, and in Doppler broadened media, where $\tau_D \ll \tau_M$, many-body corrections to the stimulated radiative operator dominate over those due to the spontaneous radiative operator.

\subsection{Extending many-body corrections to arbitrary order}
We have showed that corrections to the canonical radiative transfer equation can be parsed into terms involving the same particle and multiple radiation modes, and between different particles and a single radiation mode. We showed that the relative magnitude of many-body interactions over single-body interactions for the first correction term is $\sim \lambda^2 \ell_D \mathcal{N} (\bar{n}+1)^{-1}$, where $\ell_D = c \tau_D$ is the Doppler coherence length, and $\bar{n} = [\lambda^2/2h\nu] \int d\hat{k} \int d\nu \ \phi_{\nu} I_{\boldsymbol{k}}$ is the integrated photon-occupation number. Since single-body corrections to the radiative transfer equation have been investigated in depth elsewhere, we now proceed to consider the regime where many-body corrections to the radiative operators dominate over single-body corrections, and subsequently compute the magnitude of these corrections to arbitrary order. In the previous section, the first-order corrections to the radiative operators is worked out in detail, the result of which are, 
\begin{subequations}
\label{eq:rad_first}
\begin{align}
\sigma_1&=
-\tau_{\nu}^{\mathrm{coh}} \left[-\kappa_{\nu} I_{\boldsymbol{k}}+\epsilon_{\nu}\right] + \mathrm{s.b.}, \\
\gamma_1&=
-\tau_{\nu}^{D,\mathrm{coh}} \left[-\kappa_{\nu} I_{\boldsymbol{k}}+\epsilon_{\nu}\right] + \mathrm{s.b.},
\end{align}
\end{subequations}
where we have assumed dominant Doppler broadening, where s.b. is to indicate corrections due to single-body effects, and where we defined $\tau_{\nu}^{D,\mathrm{coh}}$ in Eq.~(\ref{eq:tau_D_coh}) as the optical depth over a Doppler coherence length. We recognize from Eq.~(\ref{eq:rad_first}) that in the first many-body correction to the radiative operators, we retrieve the usual radiative transfer equation, times a coherent optical depth. Many-body interactions involving the stimulated radiative operator, $\hat{\sigma}(t,t')$, scale with the lifetime of the transition levels, while for the spontaneous emission operator, $\hat{\gamma}(t,t')$, they scale with the Doppler time scale. Therefore, with dominant Doppler broadening, $\tau_{\nu}^{\mathrm{coh}}/\tau_{\nu}^{D,\mathrm{coh}} \gg 1$, so we can proceed by considering only the dominant corrections to the radiative transfer equation due to the stimulated radiative operator. Thus, after applying first-order many-body corrections, we find that the radiative transfer equation reads $\frac{d I_{\nu}}{ds} \simeq (1-\tau_{\nu}^{\mathrm{coh}})[- \kappa_{\nu} I_{\nu} + \epsilon_{\nu}]$.

Later on, we will explore the magnitude of $\tau_{\nu}^{\mathrm{coh}}$ for some spectral lines. For now, it will suffice to anticipate that $\tau_{\nu}^{\mathrm{coh}}$ can realistically exceed unity, which should immediately prompt us to consider even higher-order many-body correction terms to $\hat{\sigma} (t,t')$ and thus the radiative transfer equation. In order to perform such corrections, we notice, while recalling the results from Sec.~III.B.1, that the dominant next-order correction to the absorption and stimulated radiative operator,
\begin{align}
\label{eq:sig_corr}
\hat{\sigma}&(t_1,t_2) \to -\hbar^{-2} \left[[\hat{\sigma}(t_1,t_2),V(t_3)],V(t_4) \right] \nonumber \\
&= \sum_j  2\tilde{R}_{j3}(t_1,t_2) g_{\boldsymbol{k}}^2 f_j^{\boldsymbol{k}}(t_1-t_2) \nonumber \\ &\times \left[\hat{\sigma}(t_3,t_4) + \hat{\gamma}(t_3,t_4) \right],
\end{align}
returns again the stimulated radiative operator. Since corrections to $\hat{\sigma}(t,t')$ dominate over corrections to $\hat{\gamma}(t,t')$, we may apply Eq.~(\ref{eq:sig_corr}) recursively to obtain correction terms of arbitrary order. In this way, we can derive the $n$'th order many-body correction to the stimulated radiative operator,
\begin{align}
\label{eq:sig_n}
\hat{\sigma}_{n}& (t_1,t_2,\cdots,t_{2n+2}) = [g_{\boldsymbol{k}}^2]^{n} \sum_{j_1,\cdots j_{n}} \nonumber \\ &\times [2\tilde{R}_{j_13}(t_1,t_2)] [2\tilde{R}_{j_2 3}(t_3,t_4)]\cdots [2\tilde{R}_{j_{n}3}(t_{2n-1},t_{2n})] \nonumber \\ 
&\times \left[ \hat{\sigma} (t_{2n+1},t_{2n+2}) + \hat{\gamma} (t_{2n+1},t_{2n+2})  \right] \nonumber \\ &\times f_{j_1}^{\boldsymbol{k}} (t_1-t_2) \cdots f_{j_{n}}^{\boldsymbol{k}} (t_{2n-1}-t_{2n}).
\end{align}
Integrating the $n$'th correction over the appropriate time-integrals, and taking the trace with respect to the initial density operator (see Appendix~A), yields the $n$'th order many-body correction to the radiative transfer equation
\begin{align}
\label{eq:sig_n_int}
\sigma_n &=\int_0^{t_1} dt_2 \int_0^{t_2} dt_3 \cdots \int_0^{t_{2n+1}} dt_{2n+2} \ \hat{\sigma}_{n} (t_1,t_2,\cdots t_{2n+2}) \nonumber \\
&= \frac{[-\tau_{\nu}^{\mathrm{coh}}]^{n}}{n!} \left[-\kappa_{\nu} I_{\boldsymbol{k}}+\epsilon_{\nu}\right].
\end{align}
Now, we can recognize that the many-body line radiative transfer equation of Eq.~(\ref{eq:lrt_mb}) is obtained when Eq.~(\ref{eq:lrt_can}) is corrected with all the many-body interactions terms: $\sum_{n=1}^{\infty} \sigma_n$. Our results therefore show that the canonically used radiative transfer equation of Eq.~(\ref{eq:lrt_can}) is recovered only in the limit of vanishing coherent optical depth, whereas for $\tau_\nu^{\mathrm{coh}} \gtrsim 1$, collective many-body effects introduce a multiplicative correction factor whose magnitude can substantially alter line propagation.

\section{Discussion}
The results derived in the preceding sections establish that the canonical form of the line radiative transfer equation emerges only as a limiting case of a more general many-body framework. In particular, we have shown that when the coherent optical depth, $\tau_\nu^{\mathrm{coh}}$, becomes on the order of, or exceeds, unity, collective interactions between molecules or atoms modify the effective absorption and emission rates by a multiplicative factor that depends exponentially on $\tau_\nu^{\mathrm{coh}}$. In the following, we give a narrative summary of the derivation, and give an interpretation of the physics behind many-body effects in the transfer of radiation. We discuss the  physical meaning behind the coherent optical depth, $\tau_\nu^{\mathrm{coh}}$. We relate our derivation to previous derivations of the (quantum) radiative transfer equation in the literature. Afterwards, we delineate the parameter regimes in which the many-body radiative transfer equation should be applied, and we describe a laboratory experiment in which many-body effects are expected to measurably alter line radiative transfer. Finally, we examine how many-body effects influence the propagation of radiation in key astronomical transitions, focusing on the HI 21 cm line and low-$J$ CO rotational lines.

\subsection{Narrative summary of the derivation: assumptions and interpretation.}  
To facilitate the discussion, it is useful to give a narrative summary of the derivation of the many-body radiative transfer equation of the previous section, and summarize the main assumptions and their physical motivation. The starting point we take is the Dicke model \cite{dicke:54}, that describes the interaction of a multimode radiation field with a gas of $N$ randomly placed particles, using the dipole and rotating wave approximations, and assuming that the gas dimensions far exceed the wavelength dimension. We extend Dicke's formalism by including (i) the translational energy of the density operator, which introduces Doppler shifts, and (ii) finite state lifetimes and line broadening through collisions with other phases of the gas, described by impact theory \cite{baranger:58}. Following a similar operator algebra development to \citet{landi:83}, yields a general form of the radiative transfer equation (Eq.~\ref{eq:lrt_dens}), where we express the transfer of radiation in terms of operators for stimulated absorption/emission and spontaneous emission processes. In deriving Eq.~(\ref{eq:lrt_dens}) we have assumed that the photon field is localized in position-momentum space, valid when the photon mean free path $\kappa_{\nu}^{-1}=\ell_{\nu}^{\mathrm{mfp}}$ is much larger than the radiation coherence length $\ell_{\mathrm{r.c.}}$ \cite{rosato:11}. 

To recover the canonical radiative transfer equation, we assume that the system changes only marginally over the interaction time, so that the radiation field and molecular ensemble remain uncorrelated. Under this approximation we retrieve the canonical radiative transfer equation of Eq.~(\ref{eq:lrt_can}). Notably, by treating the interaction as a genuine many-body problem, by modeling the gas cloud as an $N$-body system, the correct molecular number density emerges naturally in the transfer equation. In contrast, single-particle approaches retrieve the molecular number density post-hoc by the replacement $\mathcal{V}^{-1} \to \mathcal{N}$ \cite{landi:83}.  

Proceeding, our aim is to relax the assumption of marginal system change over the interaction period. We do so by absorbing the time evolution due to radiation–matter coupling into the stimulated and spontaneous radiative operators and expanding them as operator series. This yields two classes of corrections to Eq.~(\ref{eq:lrt_can}): (i) \emph{single-body corrections}, that are associated with radiation scattering, which arise also in a single-particle framework and have been extensively studied \cite{lamb:71,landi:84,casini:14}; and (ii) \emph{many-body corrections}, which do not appear in single-particle treatments and have not been rigorously quantified before. Single-body corrections maximize the participating photon modes, but are restricted to a single molecule. Many-body corrections maximize the number of participating molecules, but involve only a single photon mode. We proceed the derivation under the assumption that many-body corrections dominate over single body corrections. In Sec.~\ref{sec:regime_mb}, we discuss the conditions under which this assumption holds.  

We then consider the many-body corrections to the stimulated and spontaneous radiative operators. The first order many-body correction to the stimulated radiative operator returns the canonical radiative transfer equation, multiplied by a factor $\tau_{\nu}^{\mathrm{coh}}$, which is the coherent optical depth. The coherent optical depth is the average number of absorption events (minus stimulated emission events) over a coherence length, which is defined by the molecular lifetime times the speed of light ($c \tau_M$). The first order many-body correction to the spontaneous radiative operator returns the canonical radiative transfer equation, multiplied by a factor $\tau_{\nu}^{\mathrm{D,coh}}$, which is the optical depth over a Doppler coherence length: $c \tau_D$.

The physical picture is as follows: when a photon field interacts with a molecule, it proceeds to travel in the forward direction. After the interaction, the photon field is correlated with the molecule it has interacted with. This correlation is present, as long as the photon field or the molecule remains unperturbed. Since we have assumed no other absorbing phases in the gas, the dominant mechanism of decorrelation of the field-molecule system is through the perturbation of the molecule. In the system we consider, the dominant mode of decorrelation of molecular states is through collisions with other gaseous particles, yielding a molecule lifetime of $\tau_M$. Thus, over a time $\tau_M$, the field-molecule system remains correlated, and if within this time, the photon field interacts with another molecule from the ensemble, then this interaction cannot be viewed in isolation from the initial interaction. Instead, in this case, radiation-matter interactions progress through a photon field-two molecule system. The likelihood that the radiation-matter interaction proceeds in such a way is $\tau_{\nu}^{\mathrm{coh}}=\kappa_{\nu} c \tau_M = \ell_{\mathrm{coh}}/\ell_{\nu}^{\mathrm{mfp}}$, relative to it proceeding through a field-single molecule interaction.

Many-body interactions through spontaneous emission work slightly differently, since there is no photon-field that `stimulates' this interaction. Two molecules in the ensemble, will emit as a single system if they can retain velocity coherence over the interaction time. The likelihood, relative to a single-molecule spontaneous emission event, that a correlated two-molecule system emits through spontaneous emission, is then $\tau_{\nu}^{\mathrm{D,\ coh}}=\kappa_{\nu} c \tau_D = \ell_{D}/\ell_{\nu}^{\mathrm{mfp}}$.

We consider the case of dominant Doppler broadening, so $\tau_{\nu}^{\mathrm{coh}} \gg \tau_{\nu}^{\mathrm{D,\ coh}}$, and we follow only the corrections to the radiative transfer equation that are the many-body corrections to the stimulated radiative operator. Corrections of this kind to higher order, return again the stimulated radiative operator. Therefore, we can apply subsequent corrections recursively to arbitrary order. We discussed the first order correction to the radiative transfer equation as being due to the radiation-matter interactions between a two molecule-field system, having a likelihood of $\tau_{\nu}^{\mathrm{coh}}$ relative to the single molecule-field interaction. The $n$'th order correction to the radiative transfer equation is the radiation-matter interactions between an $n+1$ molecule-field system. The likelihood of this interaction, relative to the single molecule-field interaction, is $[\tau_{\nu}^{\mathrm{coh}}]^n/n!$, where the $n!$ is a term that emerges due to the time-ordering of subsequent correlated interactions. Adding up corrections to the canonical radiative transfer equation in this way, to infinite order, yields the many-body radiative transfer equation of Eq.~(\ref{eq:lrt_mb}), that is the regular radiative transfer equation, multiplied by a correction factor which is the exponent of (minus) the coherent optical depth. This is the main result of this paper.

\begin{figure*}[ht!]
    \centering
    \includegraphics[%
        width=0.9\linewidth,
    ]{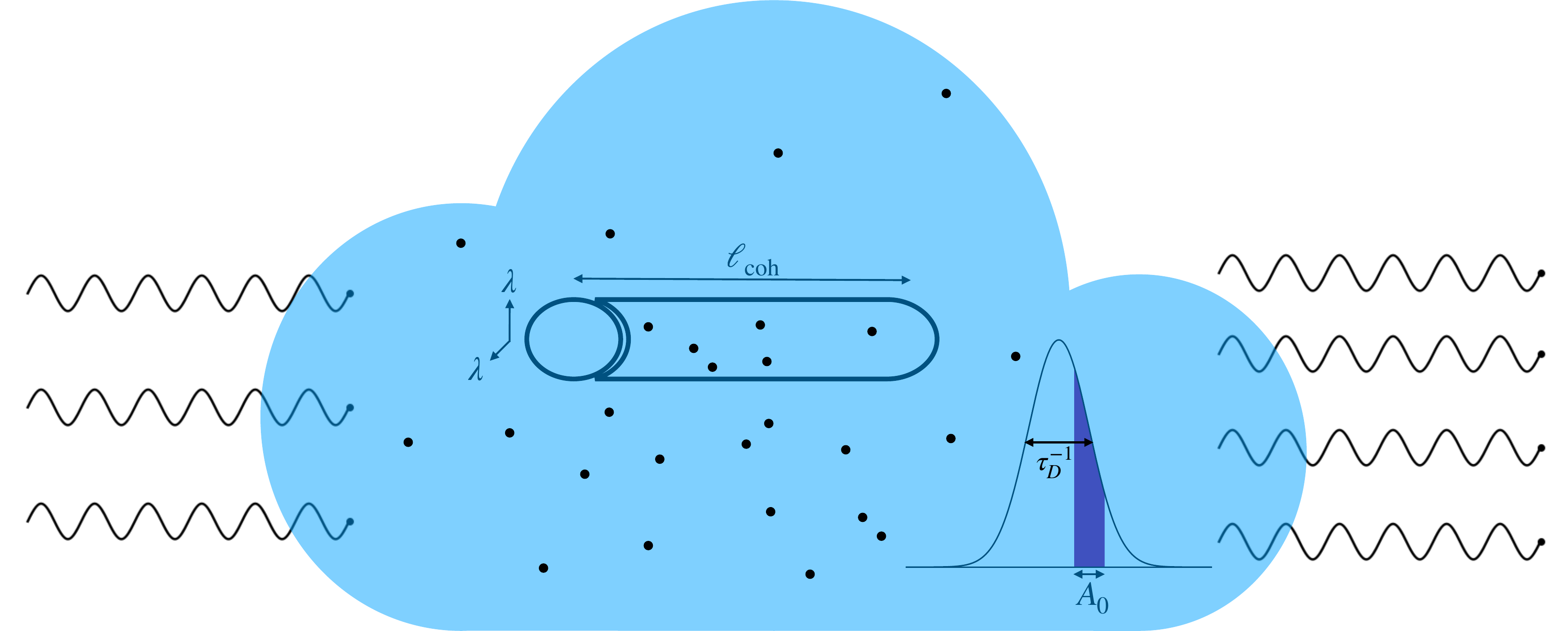}
\caption{Interpretation of the coherent optical depth, \(\tau_{\nu}^{\mathrm{coh}}\). The factor \(\tau_{\nu}^{\mathrm{coh}} \sim [\lambda^2\ell_{\mathrm{coh}} \mathcal{N} (n_- - n_+)] A_0/\Delta \nu\) can be understood as the number of molecules that interact coherently with the radiation field. Many-body interactions occur within a volume, \(\lambda^2\ell_{\mathrm{coh}}\), set by the radiation wave cross-section and coherence length. The particle density that interacts coherently is, \(\mathcal{N}(n_- - n_+)A_0 / \Delta \nu\), and is selected within a small slice of order, \(A_0 / \Delta \nu\), of the velocity distribution.}    
\label{fig:oversight}
\end{figure*}
We linger on the physical interpretation of the coherent optical depth. If small, then many-body effects are of marginal impact to the transfer of radiation, and a single-body treatment is permitted. While for large coherent optical depth, radiative transfer has to be viewed in a fundamentally different way. The coherent optical depth, we have derived to be  
$$\tau_{\nu}^{\mathrm{coh}} = \frac{h\nu}{4\pi} g_l B_{lu} \mathcal{N} \left(\frac{n_l}{g_l}-\frac{n_u}{g_u} \right) \phi_{\nu} \ell_{\mathrm{coh}},$$ 
where $g_l B_{lu}( = B_0)$ is the Einstein coefficient for stimulated absorption, $n_{u/l}(=g_{u/l}\ n_{\pm} )$ is the upper/lower level population, and $g_{u/l}$ is the upper/lower level degeneracy. The coherent optical depth admits a simple physical interpretation: it is the effective number of particles that interact collectively with the radiation field at a given frequency. A useful way to rewrite the coherent optical depth, is by using, $g_l B_{lu} = \lambda^2 g_uA_{ul} / 2 h \nu$ and $\phi_{\nu} \sim \frac{1}{\Delta \nu}$, where $\Delta \nu$ is the line width in frequency units. Dropping $\mathcal{O}(1)$ numerical factors, we then recognize
$$
\tau_{\nu}^{\mathrm{coh}} \sim \left[\lambda^2\ell_{\mathrm{coh}}\right] \left[\mathcal{N} \left(\frac{n_l}{g_l}-\frac{n_u}{g_u}\right)\right] \left[\frac{A_{ul}}{\Delta \nu
}\right]. 
$$ 
This factorization separates three elements:  
(i) a \emph{coherent interaction volume} $\lambda^2\ell_{\mathrm{coh}}$, set by the transverse area that a photon scans ($\lambda^2$) and the coherence length $l_{\mathrm{coh}}$;  
(ii) the \emph{population density} of molecules of the relevant transition, $\mathcal{N}(n_l/g_l - n_u/g_u)$; and  
(iii) a \emph{velocity–coherence fraction}
$g_u A_{ul}/\Delta \nu \ll 1$,
which (up to factors of order unity) equals the ratio of the natural line width to the Doppler width, and quantifies the fraction of the Maxwellian velocity distribution for which the Doppler detuning remains $\lesssim A_{ul}$. This defines the fraction of molecules that interact with the same photon mode. In Fig.~(\ref{fig:oversight}) we have represented the \emph{velocity–coherence fraction} as a slice of order $A_{ul}$, of the ensemble population that is distributed in velocity-space.

As such, $\tau_{\nu}^{\mathrm{coh}}$ measures the average number of molecules that are collectively coupled to the same photon-field mode. For thermal (maser) lines, when this number is large, many-body interactions reduce (boost) the effective radiative interaction rates, with the strongest suppression (enhancement) at line center where $\tau_{\nu}^{\mathrm{coh}}$ peaks.

\subsection{Relation to other derivations of the radiative transfer equation}
It is instructive to compare the derivation of the radiative transfer equation we present here with alternative radiation transport formalisms that focus on different regimes of applicability. 
\subsubsection{Single-body polarized line-formation theory}
\citet{bommier:97} derives the radiative transfer equation, including polarization modes, by modeling the interaction of a multi-mode radiation field with an isolated atom. They derive an equation analogous to our Eq.~(\ref{eq:densI_prop}), but assume that at any time the joint density matrix of the atom + radiation system can be factorized as
$
\hat\rho (t) = \hat\rho^{(M)}(t) \otimes (t)\,\hat\rho^{(R)}(t)
$
(see equation 15 of \cite{bommier:97}), thereby excluding explicit atom–radiation correlations throughout the interaction period. In contrast, we make the less stringent assumption that at $t=0$, atom-radiation correlations are not present, but explicitly allow for correlations to emerge through the interaction. \citet{bommier:97} compute corrections to the radiative transfer equation, that correspond to our `single-body' corrections in Sec.~III.B, noting that we only present the first order single body corrections. We choose to discard these, and higher-order, single-body corrections, since we focus on the regime where many-body corrections dominate. \citet{bommier:97} carries the single-body corrections to infinite order, thereby properly deriving the lifetime broadening of a line. We have verified that applying our single-body correction procedure yields the same lifetime broadening as derived by \citet{bommier:97}.

\subsubsection{Non-Boltzmann radiative transport}
In a series of papers, \citet{rosato:10,rosato:11,rosato:13} challenge the common assumption underpinning most radiative transfer formalisms, namely that the process can be cast as a Boltzmann-type equation. \citet{rosato:10} shows that this simplification only holds when the photon mean free path (i.e., the inverse absorption coefficient) is much larger than the radiative coherence length, 
$
\ell_{\mathrm{r.c.}} \sim \frac{c}{\Delta\omega_{1/2}}.
$
If this limit is not met, one must prefer the quantum radiative transfer equation, that is a transport equation for the Wigner quasi-probability distribution function of the photon field \cite{cooper:82, rosato:10}. When subsequent interactions occur within a distance of the radiative coherence length, the photon fundamental position-momentum uncertainty comes into play. In this case, photons are not well localized in position-momentum space, and different photon modes interfere to decrease the radiative coupling of the photon field to the gas \cite{rosato:11}. The formalism of \citet{rosato:10,rosato:11,rosato:13} overlaps substantially with our own set-up: in \cite{rosato:10} a radiation field interacting with many particles is modeled, and Doppler broadening is treated similarly to our approach, though collisional and lifetime broadening are included more phenomenologically. Even though \citet{rosato:10, rosato:11, rosato:13} explicitly accounts for the interaction of many particles with the radiation field in their Hamiltonian set-up, their approach does not retrieve many-body effects in their quantum radiative transfer equation, because it is assumed that the particle‐ensemble density operator and the radiation‐field density operator remain uncorrelated throughout the interaction \cite{rosato:13}. The derivation in this manuscript has shown that many-body effects emerge through correlations between different particles, that are mediated by the radiation field. Therefore, they will not arise under the assumption of uncorrelated particle-ensemble and radiation-field density operators. This is an important point, because for line radiation, typically $\ell_{\mathrm{coh}} \gtrsim \ell_{\mathrm{r.c.}}$, so the regime in which the quantum radiative transfer equation is required, is also one where many-body effects become manifest (though the converse need not hold). Many-body effects can in principle be incorporated into the quantum radiative transfer formalism, using the results of this manuscript. A full analysis of the interplay between many-body processes and photon position-momentum uncertainty in radiative transfer is beyond the present scope and will be addressed in future work.

\subsubsection{Superradiant radiative transport}
We adopt Dicke's Hamiltonian as a convenient starting point, but the physical origin of the many-body correction to the radiative transfer equation that is quantified here, differs from the mechanism usually associated with Dicke's superradiance. 

In the standard picture, the enhanced emissivity in superradiant samples is attributed to inter-particle dipole--dipole correlations. The correction we quantify, and that leads to the many-body radiative transfer equation, has a different physical origin. This can be seen by noting that the spontaneous radiative operator contains cross-correlations of the form, $\tilde{R}_{j+}(t)\tilde{R}_{l-}(t')$ , which play the role of dipole--dipole correlations. In the regime considered here, the radiative-transfer corrections associated with these terms are subdominant to those arising from the stimulated radiative operator, which couples the population operator to the radiation field, $\tilde{R}_{j3}(t,t')\,\hat{i}_{\boldsymbol{k}}$, effectively representing correlations the atomic/molecular population states and the radiation field.

This highlights a conceptual distinction between our approach, and those usually invoked when modeling Dicke's superradiance. In superradiance treatments, the relevant physics is encoded in inter-particle dipole correlations, and one therefore works with an effective master equation for the atomic/molecular ensemble density operator (e.g.~equation~3.18 of \citet{gross:82}), that is obtained after tracing out the radiation field density operator. In contrast, our formalism retains and evolves both the ensemble and radiation-field density operators, thereby keeping explicit field--ensemble correlations that are absent once the field is traced out. While such an approach has been crucial to obtain the results presented here, it is restricted to a weak radiation field regime, and is not applicable when stimulated absorption/emission rates dominate over the natural lifetime of the transition. Accordingly, while the starting point of our approach is similar to that used when modeling Dicke's superradiance, it is not intended to describe the strongly driven and/or high-gain regime in which superradiant bursts and related collective-emission phenomena are usually analyzed.

\begin{figure}[h]
  \centering
  \begin{subfigure}[t]{0.45\textwidth}
    \centering
    \includegraphics[width=\linewidth]{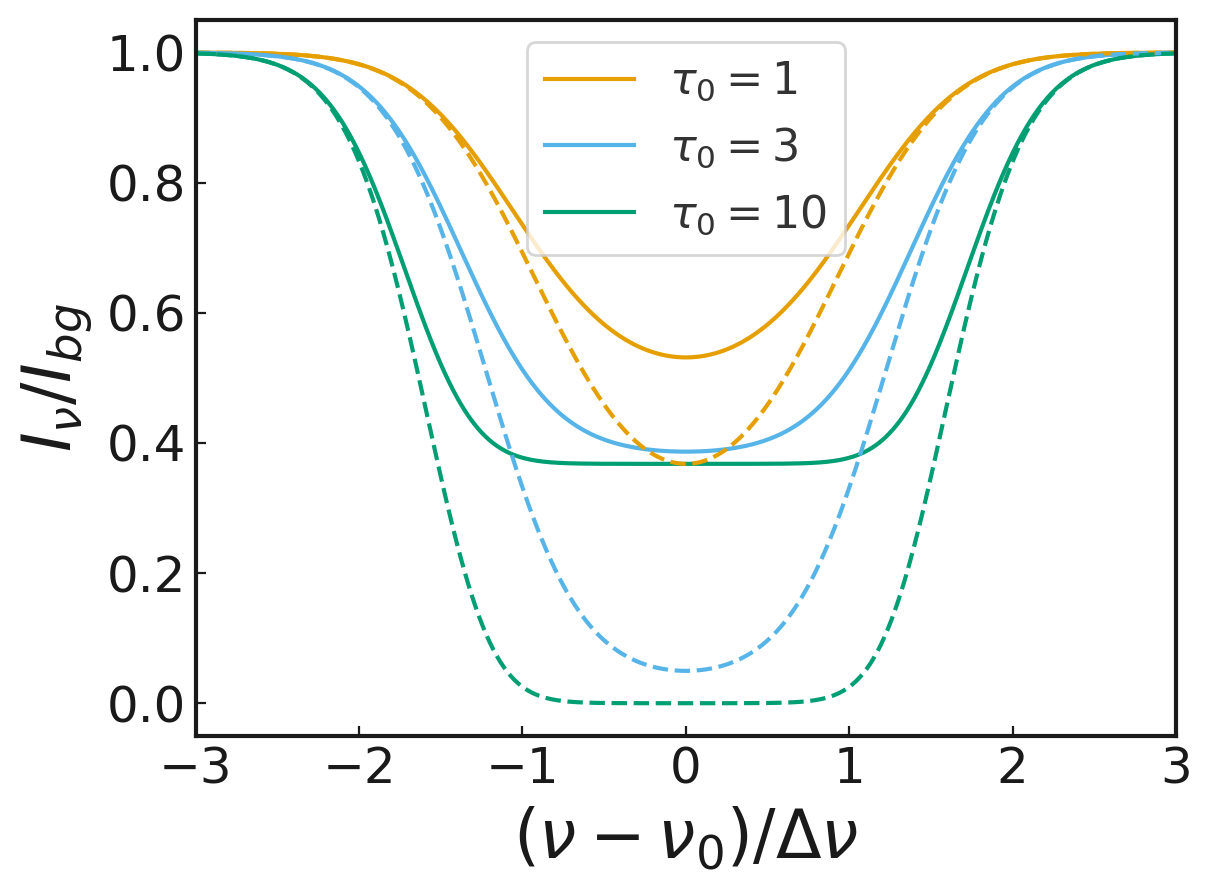}
    \caption{}
  \end{subfigure}\hfill
  \begin{subfigure}[t]{0.45\textwidth}
    \centering
    \includegraphics[width=\linewidth]{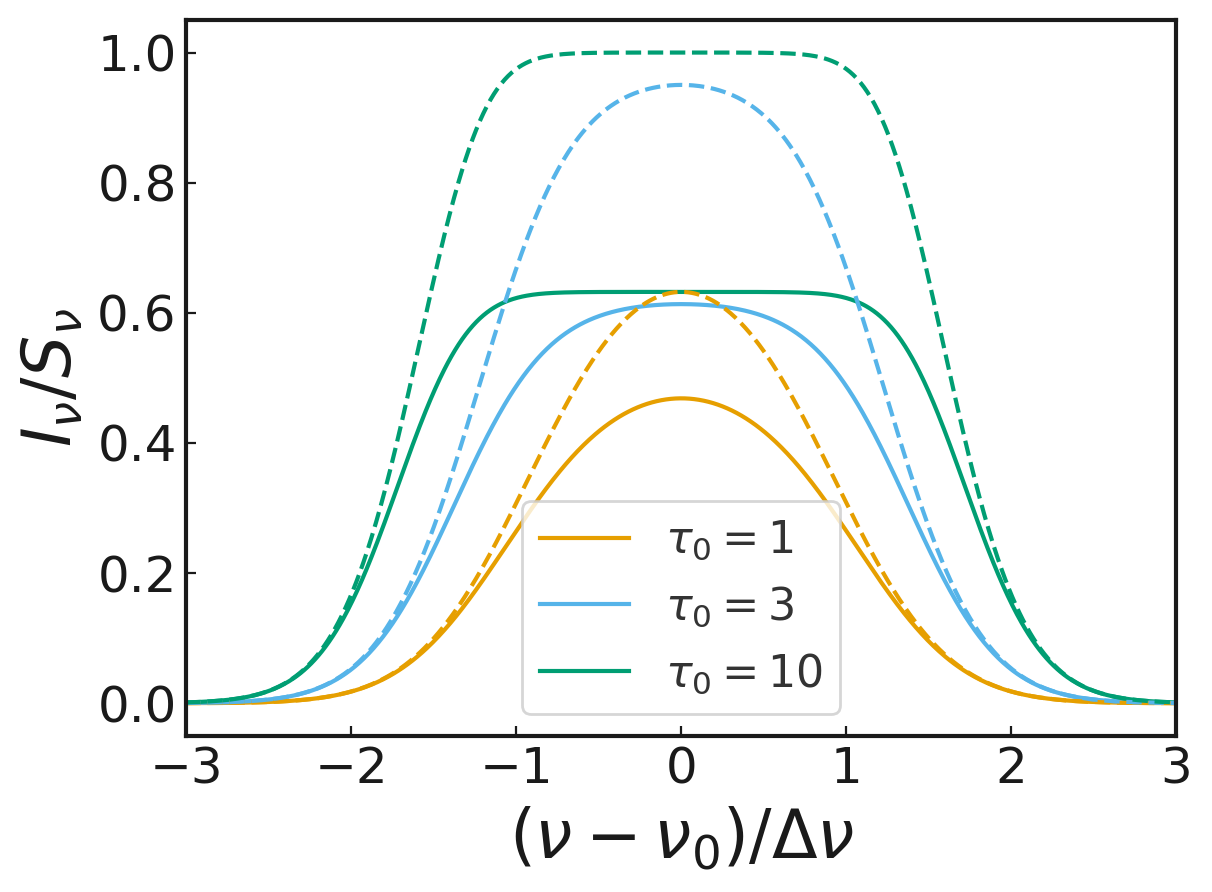}
    \caption{}
  \end{subfigure}
  \caption{Emergent spectra derived from the solution to the many-body radiative transfer equation, for the special case $\tau_{\nu}^{\mathrm{\mathrm{coh}}}>\tau_{\nu}$. We plot profiles due to dominant (a) absorption, and (b) emission, for different line-center optical depths, $\tau_0$. In dashed lines, we also plot the solution of the canonical radiative transfer equation.}
  \label{fig:coh_profiles_1}
\end{figure}

\subsection{Solutions to the many-body radiative transfer equation}
\label{sec:sol_rt}
\begin{figure}[h]
  \centering
  \begin{subfigure}[t]{0.45\textwidth}
    \centering
    \includegraphics[width=\linewidth]{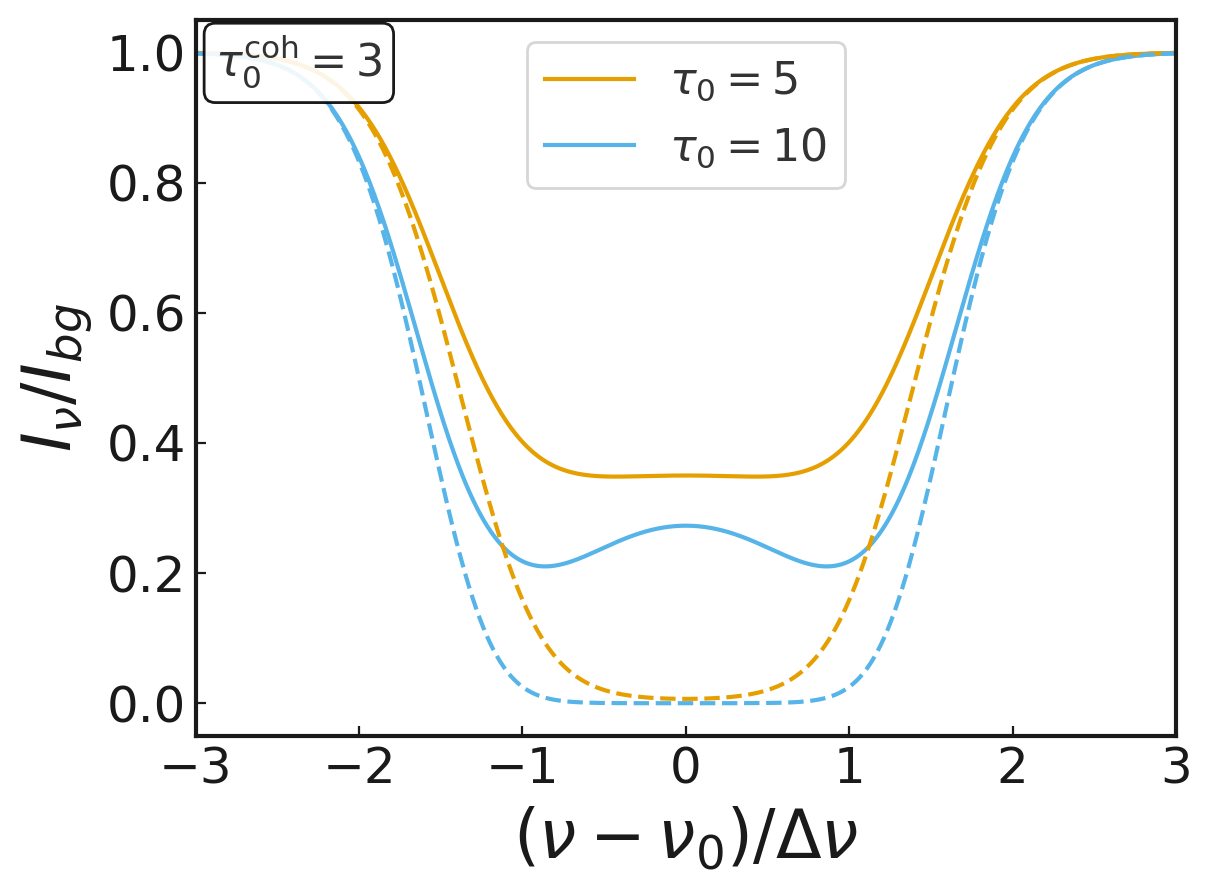}
    \caption{}
  \end{subfigure}\hfill
  \begin{subfigure}[t]{0.45\textwidth}
    \centering
    \includegraphics[width=\linewidth]{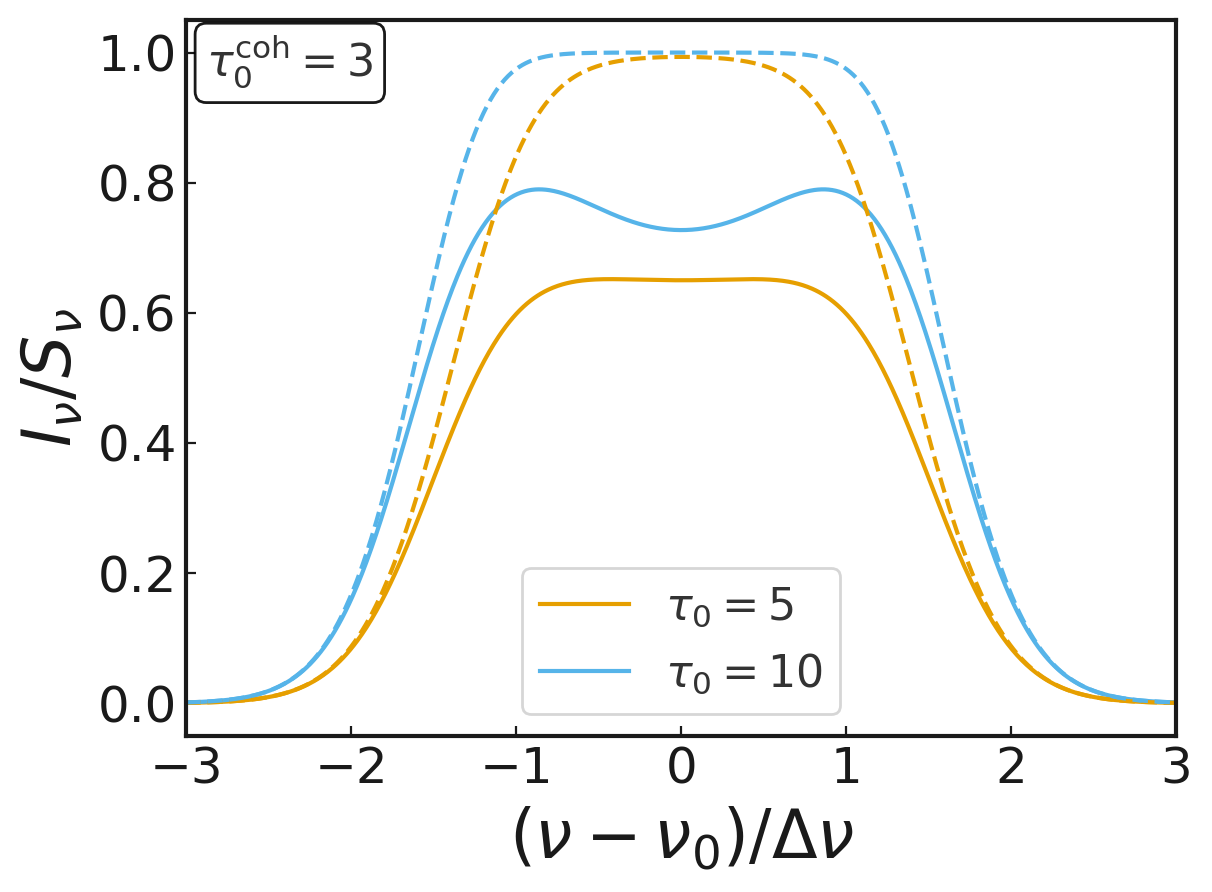}
    \caption{}
  \end{subfigure}
  \caption{Emergent spectra derived from the solution to the many-body radiative transfer equation, for the case $\tau_{\nu}^{\mathrm{\mathrm{coh}}}<\tau_{\nu}$. We plot profiles due to dominant (a) absorption, and (b) emission, for coherent optical depths at line center, $\tau_0^{\mathrm{coh}}=3$, and for different optical depths at line center $\tau_{0}$. In dashed lines, we also plot the solution of the canonical radiative transfer equation.}
  \label{fig:coh_profiles_2}
\end{figure}

We note again here the many-body radiative transfer equation
\[
\frac{d I_{\nu}}{ds} = e^{-\tau_{\nu}^{\mathrm{coh}}} \left[-\kappa_{\nu} I_{\nu} + \epsilon_{\nu}\right],
\]
for which we will find the solution in some simplified cases. We rewrite the radiative transfer equation in terms of optical depth $d\tau_{\nu} = \kappa_{\nu} ds$, define the source function $S_{\nu} = \epsilon_{\nu}/\kappa_{\nu}$, and integrate accordingly
\begin{align}
\label{eq:lrt_mb_sol}
I_{\nu} (\tau_{\nu}) = A^{-1}(\tau_{\nu})\left[I_{\nu}(0) + \int_0^{\tau_{\nu}}d\tau_{\nu}'\ e^{-\tau_{\nu}^{\mathrm{coh}}} S_{\nu} A(\tau_{\nu}')\right],
\end{align}
where 
$$A(\tau_{\nu}) = \exp \left[\int_0^{\tau_{\nu}} d\tau_{\nu}'\ e^{-\tau_{\nu}^{\mathrm{coh}}}\right].$$ Working towards the solution of Eq.~(\ref{eq:lrt_mb_sol}), we assume a homogeneous medium, through which $S_{\nu}$ is a constant. Furthermore, we put the restriction on the coherent optical depth
$$
\tau_{\nu}^{\mathrm{coh}} =
\begin{cases}
\tau_{\nu}, & \text{for } \tau_{\nu} < \tau_{\nu}^{\mathrm{coh}}, \\
\tau_{\nu}^{\mathrm{coh}}, & \text{for } \tau_{\nu} \geq \tau_{\nu}^{\mathrm{coh}} ,
\end{cases}
$$
which captures that the coherent optical depth is limited by the actual optical depth along the ray. Under these assumptions, Eq.~(\ref{eq:lrt_mb_sol}) has two types of solutions. First, for $\tau_{\nu} < \tau_{\nu}^{\mathrm{coh}}$, so that $\tau_{\nu}^{\mathrm{coh}} = \tau_{\nu}$. Then
$A(\tau_{\nu}) = \exp \left[\int_0^{\tau_{\nu}} e^{-\tau_{\nu}'} d\tau_{\nu}'\right] = \exp\!\big(1-e^{-\tau_{\nu}}\big),$ so that
\begin{align}
\label{eq:rt_sol_1}
I_{\nu}(\tau_{\nu}) 
&= S_{\nu} + (I_{\nu}(0) - S_{\nu}) \exp \left[e^{-\tau_{\nu}}- 1 \right].
\end{align}
where $I_{\nu}(0)$ is the background radiation field. In Fig.~(\ref{fig:coh_profiles_1}) we plot the solution of Eq.~(\ref{eq:rt_sol_1}) for Doppler broadened lines with $\tau_{\nu}=\tau_0 e^{-[\nu-\nu_0]^2/\Delta \nu^2}$ for $\tau_0=1,\ 3$ and $10$. We consider the case of dominant absorption ($I_{\nu}(0) \gg S_{\nu}$), and dominant emission $S_{\nu}\gg I_{\nu}(0)$, and compare to solutions that would emerge from the canonical radiative transfer equation. The solution to the many-body radiative transfer equation for a cloud, with $\tau_{\nu}^{\mathrm{coh}} < \tau_{\nu}$, is 
\begin{align}
\label{eq:rt_sol_2}
I_{\nu}(\tau_{\nu}) = S_{\nu} + \left(I_{\nu}(\tau_{\nu}^{\mathrm{coh}}) - S_{\nu}\right) \exp \left[-e^{-\tau_\nu^{\mathrm{coh}}}(\tau_{\nu}-\tau_{\nu}^{\mathrm{coh}}) \right],
\end{align}
where $I_{\nu}(\tau_\nu^{\mathrm{coh}})$ is evaluated from Eq.~(\ref{eq:rt_sol_1}). In Fig.~(\ref{fig:coh_profiles_2}) we plot the solution of Eq.~(\ref{eq:rt_sol_2}) for Doppler broadened lines with $\tau_{\nu}=\tau_0 e^{-[\nu-\nu_0]^2/\Delta \nu^2}$ for $\tau_{0}^{\mathrm{coh}}=3$ and $\tau_0=5$ and $10$. We consider the case of dominant absorption ($I_{\nu}(0) \gg S_{\nu}$), and dominant emission $S_{\nu}\gg I_{\nu}(0)$, and compare to solutions that would emerge from the canonical radiative transfer equation.

The emergent line profiles obtained from the many-body radiative transfer equation show clear departures from the standard radiative transfer solutions. In the case where the coherent optical depth equals the total optical depth, Fig.~(\ref{fig:coh_profiles_1}) shows that the absorption-dominated profiles never vanish at line center, not even for very high optical depth. Instead, they saturate at a finite floor of $I/I_{bg}=e^{-1}$. As the peak optical depth increases, the absorption profile broadens at this saturation level, in a way similar to regular radiative transfer. Conversely, in the emission-dominated regime the line intensity never reaches the source function; instead, it plateaus at a maximum value of $I/S_\nu=1-e^{-1}$. Increasing optical depth broadens the profile in a manner similar to regular radiative transfer, but leaves the saturation level unchanged. For the case where $\tau_\nu>\tau_\nu^{\mathrm{coh}}$, the emergent profiles change. Fig.~\ref{fig:coh_profiles_2} shows that in absorption, the line center drops slightly below the $e^{-1}$ floor that is obtained for $\tau_\nu=\tau_\nu^{\mathrm{coh}}$, while away from the line center the absorption deepens further. This produces symmetric profiles with absorption `shoulders' displaced from the line center. Similarly, in emission profiles for $\tau_\nu>\tau_\nu^{\mathrm{coh}}$, the line center is only slightly enhanced relative to the $\tau_\nu=\tau_\nu^{\mathrm{coh}}$ case, but stronger emission develops away from the center, producing a double-peaked, symmetric profile. The absorption/emission shoulders become more prominent with $\tau_{\nu}^{\mathrm{coh}}$ and $\tau_{\nu}/\tau_{\nu}^{\mathrm{coh}}$.

It is interesting to analyze the emergent intensity profile for lines with both coherent and regular optical depth significantly exceeding unity. We discuss the profile due to dominant emission, with a negligible background radiation field. In the high (coherent) optical depth limit, the line develops a profile a characteristic doubly peaked profile (see also Fig.~\ref{fig:coh_profiles_1}). The peaks are roughly situated at $\pm \Delta \nu \log \tau_{\nu_0}^{\mathrm{coh}}$, and they saturate to the source function when $\tau_{\nu}/\tau_{\nu_0} \gg 1$. The peak width tightens with the coherent optical depth, but will widen with the total optical depth. A convenient estimate for the peak FWHM is $\mathrm{FWHM} \sim \Delta \nu \sqrt{\frac{\log \tau_{\nu}/\tau_{\nu}^{\mathrm{coh}}}{\log \tau_{\nu}^{\mathrm{coh}}}}$. Outside of the peaks, the emission falls rapidly, while between the peaks, the spectrum will saturate to $S_{\nu}(1-1/e)$.

\begin{figure}[ht!]
    \centering
    \includegraphics[width=0.45\textwidth]{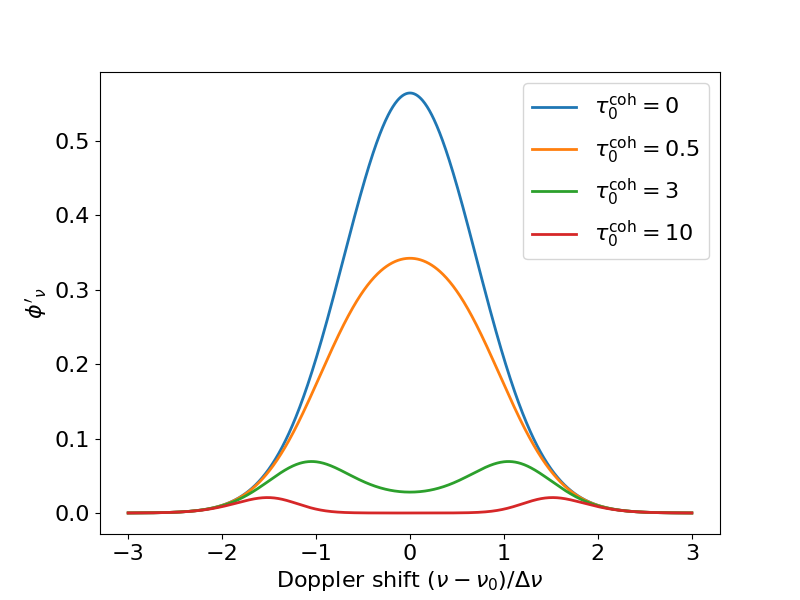} 
    \caption{Adjusted line profiles due to many-body effects. When no many-body effects are present, the integrated line profile is normalized to unity. With increasing coherent optical depth, the magnitude of the integrated line profile progressively diminishes.}
    \label{fig:coh_profiles}
\end{figure}

We continue to consider the transfer of radiation for a line with very large optical depth, and modest coherent optical depth, $\tau_{\nu}\gg \tau_{\nu}^{\mathrm{coh}}$. For such lines, we evaluate the many-body radiative transfer equation of Eq.~(\ref{eq:lrt_mb}) without considering the initial region where $\tau_{\nu}<\tau_{\nu}^{\mathrm{coh}}$. Comparing Eqs.~(\ref{eq:lrt_can}) and (\ref{eq:lrt_mb}), we recognize that the impact of many-body effects on transfer of line radiation can be neatly summarized into a transformation of the line profile: $\phi_{\nu} \to \phi_{\nu}'=e^{-\tau_{\nu}^{\mathrm{coh}}} \phi_{\nu}$. For Doppler broadened lines, the adjusted line profile is
$$\phi_{\nu}' = \frac{1}{\sqrt{\pi} \Delta \nu}e^{-[\nu-\nu_{0}]^2/\Delta \nu^2} \exp [{-\tau_{\nu_0}^{\mathrm{coh}} e^{-[\nu-\nu_0]^2/\Delta \nu^2}}],$$
where $\Delta \nu$ is the line width, and $\tau_{\nu_0}^{\mathrm{coh}}$ is the coherent optical depth at line center. In Fig.~\ref{fig:coh_profiles}, we plot the adjusted line profile for some coherent optical depths. 

We may immediately evaluate that at coherent optical depths $\tau_{\nu_0}^{\mathrm{coh}}>1$, the line profile acquires two maxima at $\nu_{\mathrm{max}}= \nu_0 \pm \Delta \nu \sqrt{\log{\tau_{\nu_0}^{\mathrm{coh}}}}$, which amount to $[\phi'_{\nu}]_{\mathrm{max}}^{\tau_{\nu_0}^{\mathrm{coh}}>1}= [e \tau_{\nu_0}^{\mathrm{coh}} \sqrt{\pi} \Delta \nu ]^{-1}$. Thus, at large coherent optical depths, the line profile acquires the appearance of a double peaked profile, with maxima that scale inversely linearly with the coherent optical depth and a separation of both peaks that scales with the logarithm of the coherent optical depth. The double peaked nature of the line profile we had also recognized from the emergent line profiles plotted in Fig.~(\ref{fig:coh_profiles_2}), when $\tau_{\nu}>\tau_{\nu}^{\mathrm{coh}}$. 

A useful analogy can be drawn between the many-body line profile and the treatment of radiative trapping developed by \citet{holstein:47}. Holstein formulated a set of equations to describe the excitation of an optically thick resonance line. In his framework, the `monochromatic transmission factor' of a gas is given by
$
e^{-\tau_{\nu}}
$
(equation 2.2 of \citet{holstein:47}). Assuming complete redistribution of velocities after each absorption event, the total transmission of radiation that is relevant for the transport of excitation is
\[
\int d\nu \, \phi_{\nu} e^{-\tau_{\nu}}
\]
(equation 2.15 of \citet{holstein:47}). The integrand bears a close resemblance to the adjusted line profile $\phi'_{\nu}$ that arises in our many-body treatment. Note that for \citet{holstein:47}, the integral of the transmission factor represents the probability that a photon escapes absorption in a resonance line. In our context, the quantity $\int d\nu\ \phi_{\nu}'$ represents a direct measure of the total reduction in radiative coupling due to many-body effects.

Evaluating the relevant integral, \citet{holstein:47} obtained
\begin{align}
\int d\nu \, \phi_{\nu}'=\int d\nu \, \phi_{\nu} e^{-\tau_{\nu}^{\mathrm{coh}}}
\simeq \frac{1}{\tau_{\nu_0}^{\mathrm{coh}} \sqrt{\pi \log \tau_{\nu_0}^{\mathrm{coh}}}},
\end{align}
valid in the limit of large coherent optical depth. We will use this result when discussing many-body effects in the propagation of CO emission in interstellar media in Sec.~\ref{sec:manybod_astro}.
%

\subsection{Regime of the many-body radiative transfer equation}
\label{sec:regime_mb}
The many-body radiative transfer equation of Eq.~(\ref{eq:lrt_mb}), that was derived in the previous sections, is valid under the following conditions:
\begin{subequations}
\label{eq:manybody_reqs}
\begin{enumerate}[label=(\alph*)]
	\item We assumed that (i) the main decorrelation process is that of the molecular state-operators, (ii) that it proceeds through collisions with other particles in the gas, (iii) and that collisions can be modeled in the impact approximation. Assumption (i) implies that decorrelation of the radiation field (operators), that can occur through interaction with another phase in the gas, that is co-spatial and where interactions occur around the same frequency as the transition frequency that we study, is relatively slow.  Assumption (ii) implies that the lifetime of our transition levels is solely determined by collisions, thus neglecting any radiative interactions with reservoir states. When discussing the lifetimes of some lines that occur in the interstellar medium, in Sec.~\ref{sec:manybod_astro}, we briefly show how to relax this approximation. Finally, regarding assumption (iii), the impact approximation is valid when the dephasing time scale and time scale between subsequent collisions that an ensemble molecule experiences, $\tau_L$, is much shorter than the time scale of a collision, $\tau_c$. An order of magnitude comparison between both time quantities permits the inequality 
\begin{align}
\label{eq:impact_regime}
\tau_c / \tau_L \sim \frac{n \sigma v_T}{v_T / \sqrt{\sigma}} \sim n \sigma^{3/2}  \ll 1,
\end{align} 
where we take $\tau_L^{-1} \sim n \sigma v_T$ as dependent on the number density, $n$, the averaged collision cross-section $\sigma$, and the thermal velocity between collision partners $v_T$, and the collision time we take as $\tau_c \sim \sqrt{\sigma} / v_T $. As a final requirement for the impact approximation to be valid: frequencies that are resonant within $\tau_c^{-1}$ have to be considered. This is obviously fulfilled for our applications.
	\item We assumed the dominant broadening mechanism to be due to Doppler broadening. This means that the time scale for Doppler decorrelation, $\tau_D$, should be much shorter than the collisional time scale. To order of magnitude, this implies that,
\begin{align}
\label{eq:doppler_regime}
\tau_D / \tau_L \sim \frac{n \sigma v_T}{\pi \sqrt{k_B T / M}/\lambda} \sim \sqrt{\frac{M}{8 \pi m}} \lambda n \sigma  \ll 1,
\end{align}
where we used $v_T = \sqrt{8 k_BT/\pi m}$, with $m$ the reduced mass of the collision complex, $T$ the gas temperature, and $k_B$ the Boltzmann constant. We furthermore note $M$ as the mass of the particle relevant to the spectral line transition. 
 	\item We assumed that many-body corrections dominate over single-body corrections. Comparing Eqs.~(\ref{eq:kv_nc_2}) and (\ref{eq:kv_c_2}), this condition is satisfied when 
\begin{align}  
\label{eq:manybody_regime}
\frac{\tau^{\mathrm{coh}}_{\nu}}{A_{0} (\bar{n}+1) \tau_M/2} = \frac{\lambda^2}{4\pi(\bar{n}+1)} g_u \mathcal{N} \left(\frac{n_l}{g_l} - \frac{n_u}{g_u} \right) c \phi_{\nu} > 1,
\end{align}
where we recall that $\lambda$ is the wavelength, $\bar{n}$ is the integrated photon occupancy number, $g_{u/l}$ are the upper/lower state degeneracies, $\mathcal{N}n_{u/l}$ are the upper/lower transition states number densities, and $\phi_{\nu}$ is the line profile. 
  \item The final constraint comes from the steady-state form of the radiative transfer equation. In order for steady-state to be operative, we require the rates of absorption and stimulated emission to be slower than the decay rate (inverse lifetime) of the transition states,
\begin{align}
\label{eq:nolaser_regime}
A_{ul} \bar{n} \tau_M < 1,
\end{align}  
where we recall that $A_{ul}$ is the Einstein A-coefficient. This condition is commonly fulfilled, except when strong laser radiation is resonant with the transition under investigation. In such cases, a Maxwell-Bloch formalism has to be set up when modeling the transfer of radiation, and its impact on the transition states \cite{lehmberg:70, rajabi:20}.
\end{enumerate}
Beyond the assumptions that must hold in specific parameter regimes, the manifestation of many-body effects ultimately requires a sufficient coherent optical depth
\begin{align}
\label{eq:tau_coh_app}
 \tau_{\nu}^{\mathrm{coh}} &=    \frac{h\nu}{4\pi} g_l B_{lu} \mathcal{N} \left(\frac{n_l}{g_l}-\frac{n_u}{g_u} \right) \phi_{\nu} [c\tau_M] \nonumber \\
            &=    \frac{\lambda^2}{8\pi} g_u A_{ul} \mathcal{N} \left(\frac{n_l}{g_l}-\frac{n_u}{g_u} \right) \phi_{\nu} [c\tau_M] \gtrsim 1.
\end{align}
Furthermore, since we have the obvious requirement that the regular optical depth, $\tau_{\nu} \gtrsim \tau_{\nu}^{\mathrm{coh}}$, we require 
\begin{align}
\label{eq:tau_app}
 \tau_{\nu} 
            &=    \frac{\lambda^2}{8\pi} g_u A_{ul} \mathcal{N} \left(\frac{n_l}{g_l}-\frac{n_u}{g_u} \right) \phi_{\nu} \ell \gtrsim 1,
\end{align}
\end{subequations}
assuming that the excitation conditions and density along the line-of-sight remain invariant for the cloud/experiment/sample size $\ell$. 

\subsection{Experimental features to accommodate many-body effects}
Given the criteria specified in Eqs.~(\ref{eq:manybody_reqs}), we can now delineate the experimental conditions necessary for many-body effects to manifest in the radiative transfer equation. In the following, we consider numerical estimates of Eqs.~(\ref{eq:manybody_reqs}) in a typical transition occurring in the microwave, mid-infrared (IR) and visible ranges of the spectrum. These order-of-magnitude estimates will serve as a basis for proposing a concrete experimental design later on.

We begin with the requirement that we perform our experiment in a density that is appropriate to treat collisions in the impact approximation. Applying the condition of Eq.~(\ref{eq:impact_regime}), puts the number density $n \ll \sigma^{-3/2}$, which, when the collisional cross-section is normalized to a standard value, puts a limit on the number density of
\begin{subequations}
\begin{align}
n \ll 10^{22.5} \mathrm{cm}^{-3} \left[\frac{\sigma}{10^{-15}\ \mathrm{cm}^{2}} \right]^{-3/2}.
\end{align}
An additional limit to the number density has to be put from the condition of dominant Doppler broadening, that we formulated in Eq.~(\ref{eq:doppler_regime}). Normalizing the wavelength to mid-IR experiments, we have the requirement
\begin{align}
n \ll 5\times 10^{18} \mathrm{cm}^{-3} \left[\frac{\lambda}{10 \ \mathrm{\mu m}} \right]^{-1} \left[\frac{\sigma}{10^{-15}\ \mathrm{cm}^{2}} \right]^{-1} \sqrt{\frac{m}{M}},
\end{align} 
which puts a more much more stringent limit on the number density than the conditions for the impact approximation. Note that for microwave experiments, the required densities should lay 3-4 orders lower than in the IR, while experiments in the visible range would have a bit less stringent constraints on the number densities. 

Since we want many-body corrections to the radiative transfer equation to dominate over single body corrections, we require the condition of Eq.~(\ref{eq:manybody_regime}) to be fulfilled. The most important variable defining this inequality lies in the relative population difference
\begin{align}
\Delta n = \mathcal{N} \left(\frac{n_l}{g_l} - \frac{n_u}{g_u}\right).
\end{align}
For a thermalized molecule, we can relate $\mathcal{N} = n x_{\mathrm{ab}} e^{-E_l / k_BT} / Q$ to the number density, through the abundance $x_{\mathrm{ab}}$, partition function $Q$, and lower level energy $E_l$. Evaluating Eq.~(\ref{eq:manybody_regime}), using standard room-temperature experimental values for a mid-IR experiments, yields the requirement for the population difference 
\begin{align}
\Delta n &> 1 \times 10^{5}\ \mathrm{cm}^{-3} \left[\frac{\lambda}{10 \ \mathrm{\mu m}} \right]^{-3} \left[\frac{T}{293 \ \mathrm{K}} \right]^{1/2} \nonumber \\ &\times\left[\frac{M}{10 \ \mathrm{m_H}} \right]^{-1/2} \left(\bar{n}+1 \right).
\end{align} 
Note that the requirement for the population difference is much more stringent for visible lines: for a $\lambda = 500$ nm line, for instance, using the same parameters, would require a population difference in excess of $1.6 \times 10^9$ cm$^{-3}$. 

In addition, we require a sample that has significant coherent and regular optical depth ($\tau_{\nu}^{\mathrm{coh}} > 1$, $\tau_{\nu} > 1$). Evaluating the coherent optical depth with Eq.~(\ref{eq:tau_coh_app}), whilst using standard room temperature experimental values, and putting $\tau_M^{-1} = n \sigma v_T$, puts constraints on the relative population difference
\begin{align}
\frac{\Delta n}{n} &> 1.5 \times 10^{-8} \left[\frac{\mu}{1 \ \mathrm{D}} \right]^{-2} \nonumber \\ &\times\left[\frac{\sigma}{10^{-15} \ \mathrm{cm}^2} \right] \left[\frac{T}{293 \ \mathrm{K}} \right] \left[\frac{M m}{10 \ \mathrm{m_H}} \right]^{-1},
\end{align}
where we evaluated $\phi_{\nu}$ at line center assuming a Doppler profile at kinetic temperature $T$, and where $\mu$ is the transition dipole moment. The total optical depth constrains the total population difference,
\begin{align}
\Delta n &> 1.5 \times 10^{11} \ \mathrm{cm}^{-3}\ \left[\frac{\mu}{1 \ \mathrm{D}} \right]^{-2}  \left[\frac{T}{293 \ \mathrm{K}} \right]^{1/2} \nonumber \\ &\times \left[\frac{m}{10 \ \mathrm{m_H}} \right]^{-1/2} \left[\frac{\ell}{50 \ \mathrm{cm}} \right]^{-1},
\end{align}
where we used $\ell$ for the dimension of the absorption cell.

Finally, the condition of Eq.~(\ref{eq:nolaser_regime}), puts a constraint on the probing light intensity. In case $A_{ul} \bar{n} \tau_M$ exceeds $1$, then the steady-state approximation used is not appropriate, and we have to a time-dependent formalism appropriate for maser/laser transitions. Rewriting Eq.~(\ref{eq:nolaser_regime}), using standard experimental values, and assuming Rayleigh-Jeans scaling for the light intensity (i.e.~$\bar{n} = k_B T_B/ h \nu$), yields the constraint on the probing light intensity (in temperature units)
\begin{align}
T_B &< 4 \times 10^7 \ \mathrm{K} \left[\frac{\lambda}{10 \ \mathrm{\mu m}} \right]^2 \left[\frac{\mu}{1 \ \mathrm{D}} \right]^{-2} \left[\frac{n}{10^{17}\ \mathrm{cm}^{-3}} \right]\nonumber \\ &\times  \left[\frac{\sigma}{10^{-15}\ \mathrm{cm}^{2}} \right] \left[\frac{T}{293\ \mathrm{K}} \right]^{1/2} \left[\frac{m}{10 \ \mathrm{m}_H} \right]^{-1/2},
\end{align}
or
\begin{align}
\bar{n} &< 3 \times 10^4\left[\frac{\lambda}{10 \ \mathrm{\mu m}} \right]^3 \left[\frac{\mu}{1 \ \mathrm{D}} \right]^{-2} \left[\frac{n}{10^{17}\ \mathrm{cm}^{-3}} \right] \nonumber \\ &\times  \left[\frac{\sigma}{10^{-15}\ \mathrm{cm}^{2}} \right] \left[\frac{T}{293\ \mathrm{K}} \right]^{1/2} \left[\frac{m}{10 \ \mathrm{m}_H} \right]^{-1/2},    
\end{align}
in terms of the dimensionless photon occupation number.
\end{subequations}
\subsubsection{An experiment to elicit many-body effects in the transfer of radiation}
Having identified the physical conditions under which many-body effects are expected to occur, we next outline an experimental setup that satisfies these criteria while remaining experimentally tractable. We choose a room-temperature rovibrational line in a simple linear molecule with well-characterized parameters and no (hyper)fine structure. We find that Carbonyl sulfide (OCS) meets these criteria.

The experimental set-up should ensure dominant Doppler broadening, so Eq.~(\ref{eq:doppler_regime}) should be fulfilled. For the least stringent constraints for a rovibrational transition on the experiment number density, we aim for the CO-stretch in OCS, that occurs at $2062$ cm$^{-1}$. In a pure OCS gas, the FWHM of Doppler broadening, at $296$ K, and self-broadening match, 
$2062 \ \mathrm{cm}^{-1} \sqrt{8 \mathrm{ln} 2 k_BT / mc^2} = 2 \gamma_{\mathrm{self}} k_BT\ n_{\mathrm{match}}$,
at $n_{\mathrm{match}}\sim 10^{17}$ cm$^{-3}$, where we have taken $\gamma_{\mathrm{self}}=0.3$ cm$^{-1}$ atm$^{-1}$ from the HITRAN database \cite{hitran:16}. An experiment at $n = 10^{16}$ cm$^{-3}$, that corresponds to $0.3$ Torr at 296 K, should thus yield marginal influence of pressure broadening and fulfill condition Eq.~(\ref{eq:doppler_regime}). 

Many-body effects only manifest at significant (coherent) optical depth. Focusing on the R(1) transition of the rovibrational manifold around $2062 \ \mathrm{cm}^{-1}$, we estimate the total and coherent optical depth. We first compute the opacity through the spectral line intensity from the HITRAN database, which at line center is $\kappa_{\nu_0} = S(T) n g(\bar{\nu}_0) = 0.26\ \mathrm{cm}^{-1}$, adopting the spectral line intensity, $S(296\ \mathrm{K}) = 8.897\times 10^{-20}$ cm / molecule cm$^{-2}$ \cite{hitran:16}, and where $g(\bar{\nu})$ is the line-profile in wave number space with line center at $\bar \nu_0$.

We let $\ell$ be the dimension of the absorption cell, while $\ell_{\mathrm{coh}} = c \tau_M$, is the coherence length. We compute from the self-broadening coefficients, that $\ell_{\mathrm{coh}} = 13$ m, at $n = 10^{16} \ \mathrm{cm}^{-3}$. We don't anticipate an experiment where $\ell > \ell_{\mathrm{coh}}$, so the experimental dimension determines the coherent optical depth $\tau_{\nu}^{\mathrm{coh}} \to \tau_{\nu} $. The emergent absorption or emission spectra from such an experiment we discussed in Sec.~\ref{sec:sol_rt} and Fig.~(\ref{fig:coh_profiles_1}).

We also require for our experiment that many-body corrections to the radiative transfer equation dominate. In order to validate this, we evaluate the condition of Eq.~(\ref{eq:manybody_regime}), computing that $\tau_{\nu_0}^{\mathrm{coh}}/A_0 \tau_M (\bar{n}+1) = 3 \times 10^{7} /(\bar{n}+1)$. This comfortably ensures dominant many-body corrections as long as a relatively weak probing field, of $\bar{n} < 1/A_0\tau_M \sim 2 \times 10^5 $ is used.

Taken together, our estimates indicate that an absorption experiment on the R(1) transition of the CO stretch in pure OCS gas ($n=10^{16}$ cm$^{-3}$, $296$ K, 1 m path length) would produce a line-center (coherent) optical depth of $\kappa_{\nu_0}\ell = 26$. This proposed experiment would fulfill all the requirements formulated in Eq.~(\ref{eq:manybody_reqs}), and should readily show many-body effects manifest in the transfer of radiation. Beyond demonstrating interparticle coherence under room-temperature conditions, such an experiment would also provide a relatively inexpensive route to determining key spectroscopic parameters, including the intrinsic molecular lifetime.

\subsection{Many-body effects for naturally occurring astronomical lines}
\label{sec:manybod_astro}
While the preceding subsections have focused on the physical interpretation, validity conditions, and possible laboratory realizations of many-body effects in line radiative transfer, the same formalism applies directly to natural environments. In particular, the very tenuous gas found in the interstellar medium naturally fulfills the conditions of Eqs.~(\ref{eq:manybody_reqs}), and many-body effects could manifest here, provided sufficient coherent optical depths: $\tau_\nu^{\mathrm{coh}} \gtrsim 1$. In the following, we explore the strong, low-frequency transitions of the ubiquitous HI 21\,cm line and the low-$J$ rotational transitions of linear molecules in general, and CO in particular. 


\subsubsection{Linear molecules in the ISM}
We consider the rotational transition, $J \to J-1$, of a linear molecule. We use the standard relations,
\begin{subequations}
\begin{align}
\label{eq:lin_freq_A}
    \nu_0 &=  2 B_{\mathrm{rot}} J, \\
    g_J A_{J,J-1} &=  \frac{512 \pi^4 B_{\mathrm{rot}}^3 \mu^2}{3 h c^3} J^4, 
\end{align}
where $\nu_0$ and $A_{J,J+1}$ are the frequency and Einstein $A$-coefficient of the $J\to J-1$ transition, $g_J=2J+1$ is the degeneracy, $B_{\mathrm{rot}}$ is the rotational constant and $\mu$ is the molecular dipole moment. Eqs.~(\ref{eq:lin_freq_A}) yield solid approximations of the molecular properties that are relevant to the opacity. We furthermore assume LTE, under which,
\begin{align}
\mathcal{N}\left(\frac{n_l}{g_l}-\frac{n_u}{g_u} \right) &= n
\frac{x_{\mathrm{ab}}}{Q} \left(\frac{n_l}{g_l}-\frac{n_u}{g_u} \right)  \nonumber \\
    &\approx n {x_{\mathrm{ab}}} 2J \left[\frac{hB_{\mathrm{rot}}}{k_BT}\right]^2,
\end{align}
where we called $Q$ the partition function, $x_{\mathrm{ab}}$ is the molecular abundance relative to H$_2$, and we let $T$ be the temperature, and where the approximation is valid as long as $h B_{\mathrm{rot}} / k_B T \ll 1$. The line profile of an astronomical gas volume of molecules we factorize as,
\begin{align}
\label{eq:phinu_app}
\phi_{\nu} &= \frac{1}{\sqrt{\pi} \Delta \nu} e^{-[\nu-\nu_0]^2/\Delta \nu^2} \nonumber \\ &\simeq \frac{1}{b+1} \left[\frac{M}{2\pi k_BT}\right]^{1/2}  e^{-[\nu-\nu_0]^2/\Delta \nu^2},
\end{align}
where we have introduced the variable $b$ that quantifies the degree of non-thermal broadening, for instance due to larger-scale turbulent motions, that is important in astrophysical gases. We note that we defined the total line width over a coherence length as $\Delta \nu = \sqrt{\frac{2k_BT}{M}}(1+b)$, where we require $b>0$, and that $b=0$ corresponds to the absence of non-thermal gas motions over a coherence length. 

The state lifetimes of linear molecules in the ISM are somewhat more complicated. In the rest of the manuscript, we have modeled the lifetime as the consequence of only collisions, but for an accurate estimate for the lifetime of molecular states in the ISM, we should also include pathways of radiative relaxation to other states. The lifetime due to collisions we can approximately model as,
\begin{align}
\label{eq:tau_M_col}
    \frac{1}{\tau_M^{\mathrm{col}}} = n \sigma v_{\mathrm{T}},
\end{align}
where we recall that $n$ is the number density, $\sigma$ the (averaged) collisional cross-section and $v_T$ the thermal relative velocity between the collision partners. At the low densities of the ISM, the lifetime of the states can be significantly impact by radiative relaxation. This relaxation to other states differs for the upper and lower transition state. For the lower state of the transition, $J-1$, assuming a thermalized radiation field at $T_{\mathrm{rad}}$, radiative relaxation occurs at a rate
\begin{align}
    \frac{1}{\tau_M^{\mathrm{rad,\ l}}} &= A_{J-1,J-2} \left( [e^{2hB_{\mathrm{rot}}(J-1)/k_BT_{\mathrm{rad}}} - 1]^{-1} +1 \right) \nonumber \\ &\simeq A_{J-1,J-2} \left(\frac{k_BT_{\mathrm{rad}}}{2hB_{\mathrm{rot}}(J-1)} + 1 \right),
\end{align}
and includes stimulated and spontaneous relaxation to the state $J-2$, while for the upper state, $J$, radiative relaxation occurs through absorption to the $J+1$ level at a rate,
\begin{align}
    \frac{1}{\tau_M^{\mathrm{rad,\ u}}}&= A_{J+1,J} \left( [e^{2hB_{\mathrm{rot}}(J+1)/k_B T_{\mathrm{rad}}} - 1]^{-1} \right) \nonumber \\ &\simeq A_{J+1,J} \left(\frac{k_B T_{\mathrm{rad}}}{2h B_{\mathrm{rot}}(J+1)} \right).
\end{align}
In most cases, $\tau_M^{\mathrm{rad,\ u}} \ll \tau_M^{\mathrm{rad,\ l}}$, due to the $\nu^3$ dependence of the Einstein A-coefficient, so that $1/\tau_{M}^{\mathrm{rad}}= 1/\tau_{M}^{\mathrm{rad,\ l}} + 1/\tau_{M}^{\mathrm{rad,\ u}} \approx 1/\tau_{M}^{\mathrm{rad,\ u}}$. When molecules are excited at densities in excess of the transition critical density, then $\tau_M \sim \tau_M^{\mathrm{col}} $, but in other cases, radiative relaxation plays an important role in determining the lifetime of the molecules. We define the factor
\begin{align}
\mathcal{R} = \frac{\tau_M^{\mathrm{col}}}{\tau_M^{\mathrm{rad}}} \approx \frac{\tau_M^{\mathrm{col}}}{\tau_M^{\mathrm{rad,\ u}}},
\end{align}
\end{subequations}
that characterizes the relative importance of radiative relaxation and let $\tau_M = \tau_M^{\mathrm{col}}/(1+\mathcal{R})$. We use the above results and approximations, to evaluate the coherent optical depth of a general diatomic line at line center, to be
\begin{align}
 &[\tau_{\nu_0}^{\mathrm{coh}}]_{J,J-1} \nonumber \\ &= \frac{4 \pi^3 \mu^2 c}{3 h} \frac{1}{1+b} \frac{1}{1+R}x_{\mathrm{ab}} J^2 \left(\frac{h B_{\mathrm{rot}}}{k_BT} \right)^2 \frac{\sqrt{M m}}{k_BT} \sigma^{-1} \nonumber \\
 &=192\ J^2 \left[\frac{1}{1+b} \right] \left[\frac{1}{1+\mathcal{R}}  \right] \left[ \frac{T}{10 \ \mathrm{K}}\right]^{-3}\nonumber \left[ \frac{B_{\mathrm{rot}}}{56 \ \mathrm{GHz}}\right]^{2} \\ &\times  \left[\frac{\mu}{0.122 \ \mathrm{D}}\right]^2 \left[\frac{x_{\mathrm{ab}}}{10^{-4}}\right] \left[\frac{\sigma}{10^{-15}\ \mathrm{cm}^{-2}}\right]^{-1} \left[ \frac{\sqrt{M m}}{56 \ \mathrm{m_H}}\right],
\end{align}
where we normalized the molecular constants to those of CO, and $\mathrm{m_H}$ is the hydrogen mass. Estimating the $b$ and $\mathcal{R}$ coefficients, we specialize to CO, and adopt typical molecular ISM gas of number density of the order $10^3$ cm$^{-3}$ and temperature $T=10$ K. Starting with an estimate for $\mathcal{R}$, we assume the lifetime is determined by absorption from the upper level, and that the radiation field is thermalized with the gas temperature. Adopting the approximation that $k_BT \gg h \nu$, we obtain 
$\frac{1}{\tau_M^{\mathrm{rad}}} \sim \frac{256 \pi^4 B_{\mathrm{rot}}^2 \mu^2}{3 h^2 c^3} \frac{J^4}{(2J+1)(J+1)} k_BT_{\mathrm{rad}} \approx7.3 \times 10^{-8} \ \mathrm{s}^{-1},$ 
yielding an $\mathcal{R} \sim 2.3$. With the lifetime determining the coherence length, $\ell_{\mathrm{coh}} = c \tau_M \approx 0.07 \ \mathrm{pc},$ which is about twice the sonic scale of interstellar medium gas. At the sonic scale, the velocity dispersion is $\Delta v \approx \sqrt{k_BT/m_{\mathrm{H}}}$, so the factor $b \sim \sqrt{M/m_\mathrm{H_2}}=3.7$ ($m_\mathrm{H_2}$ is the molecular hydrogen mass). Thus, when the corrections for the lifetime and the non-thermal broadening are taken into account, we retrieve a coherent optical depth at the line center for typical interstellar CO lines of $\tau_{\nu_0}^{\mathrm{coh}} \sim 12$.

At higher densities, the gas gets increasingly thermalized and radiative interactions become less important to the lifetime, so $\mathcal{R}$ falls. At the same time, with $\tau_M \sim \tau_M^{\mathrm{col}} \propto n^{-1}$, the lifetime drops with the density, so the coherence length $\ell_{\mathrm{coh}}$ drops with it. At lower coherence lengths, non-thermal broadening effects become less pronounced; if one assumes a Kolmogorov scaling, $\Delta v_{\mathrm{n.t.}} \propto \ell_{\mathrm{coh}}^{1/3}$. Thus, the factor $b$ also drops with the density. We therefore expect many-body effects to be more prominent for denser gases. On the other hand, for a molecule such as CO, grain adsorption becomes more pronounced at higher densities \cite{bergin:07}, thus lowering its overall abundance and therewith the coherent optical depth.

Now we turn to evaluate the dominance of many-body effects over single-body effects.
Specializing Eq.~(\ref{eq:manybody_regime}) to a rotational transition of a linear molecule, where the radiation field is equilibrated to the kinetic temperature, $T$, we find
\begin{align}    
&\frac{\tau^{\mathrm{coh}}_{\nu}}{A_{0} (\bar{n}+1) \tau_M/2} \approx \frac{\lambda^3}{4\pi} g_u \frac{n x_{\mathrm{ab}}}{Q}\left(\frac{h\nu}{k_BT} \right)^2 \sqrt{\frac{M}{2\pi k_BT}}  \nonumber \\ 
&\approx \frac{c^4}{8\pi B_{\mathrm{rot}}^3} \frac{2J+1}{J} n x_{\mathrm{ab}} \left(\frac{hB_{\mathrm{rot}}}{k_BT} \right)^3 \sqrt{\frac{M}{2\pi k_BT}} \nonumber \\
&= 28\  \frac{2J+1}{J} \left(\frac{n x_{\mathrm{ab}}}{0.1\ \mathrm{cm}^{-3}}\right) \left(\frac{T}{10 \ \mathrm{K}} \right)^{-3.5} \left(\frac{M}{28 \ \mathrm{m_H}} \right)^{0.5}  .
\end{align}
For the $J=1\to0$ transition of CO, this yields a ratio of $86$, thus ensuring the dominance of many-body effects over single-body effects. Yet, we should note that at densities of about two orders of magnitude less, the dominance of many-body terms is not guaranteed. 

Having established that the coherent optical depth of CO rotational line transitions can be $\gtrsim 1$, we investigate the impact of many-body effects on the transfer of radiation further. CO rotational lines usually have large optical depth. We therefore capture the impact of many-body effects on the radiative transfer in the adjusted line profile: $\phi_{\nu} \to \phi_{\nu}'=e^{-\tau_{\nu}^{\mathrm{coh}}} \phi_{\nu}$. As we have discussed in Sec.~\ref{sec:sol_rt}, the line profile acquires the appearance of a double peaked profile for large coherent optical depths, with maxima that scale inversely linearly with the coherent optical depth and a separation of both peaks that scales with the logarithm of the coherent optical depth. A natural objection is that the double-peaked profile that characterizes $\phi'_{\nu}$ at high coherent optical depths, cannot be commonly recognized in CO emission or absorption signals from interstellar media. To address this concern, we note that lines broaden---either by systemic kinematics or through effects of turbulence---over many coherence lengths. Therefore, their apparent width far exceeds the line width over a coherence length, which the coherent optical depths are based on. Modest turbulent broadening of the order $\Delta \nu_{\mathrm{turb}} \gtrsim 2\Delta \nu \sqrt{\log{\tau_{0}^{\mathrm{coh}}}}$, would render the unconvoluted double-peaked profile of $\phi'_{\nu}$ indistinguishable from a Gaussian line profile. For $\tau_{0}^{\mathrm{coh}}=10,\ 100, \ 1000$, this would correspond to a turbulent broadening of $\Delta \nu_{\mathrm{turb}}/\Delta \nu \sim 3,\ 4,\ 5$ times the line width over a coherence length. Such broadening is very likely in interstellar media. Consequently, the profile of $\phi'_{\nu}$, that is double-peaked due to many-body effects, would be effectively cloaked by turbulent broadening. The impact of many-body effects that remains then, is the lowering of the emissivity and absorbance, or better said, the effective optical depth, of the line by a factor $f(\tau_{\nu_0}^{\mathrm{coh}}) = \int_{0}^{\infty} d\nu\ \phi_{\nu}' \xrightarrow{\tau_{\nu_0}^{\mathrm{coh}} \gg 1} \left[\tau_{\nu_0}^{\mathrm{coh}}\sqrt{\pi \log{\tau_{\nu_0}^{\mathrm{coh}}}}\right]^{-1}.$ Thus, for lines that are significantly broadened with respect to the coherent length scale, this means that Eq.~(\ref{eq:lrt_can}) can be used, but with adjusted radiative transfer coefficients: $\kappa_{\nu} \to \kappa_{\nu} f(\tau_{\nu_0}^{\mathrm{coh}}) $ and $\epsilon_{\nu} \to \epsilon_{\nu}f(\tau_{\nu_0}^{\mathrm{coh}})$.

In summary, for linear molecules in cold interstellar gas the scaling derived above for $[\tau_{\nu_0}^{\mathrm{coh}}]_{J,J-1}$ shows that $\tau_{\nu_0}^{\mathrm{coh}}\!\propto\! J^{2}\,T^{-3}\,x_{\mathrm{ab}}\,\sigma^{-1}(1+b)^{-1}(1+\mathcal{R})^{-1}$, where we compute that low-$J$ transitions of CO at $T\!\sim\!10$~K attain $\tau_{\nu_0}^{\mathrm{coh}}\!\gtrsim\!1$. In this regime, many-body effects are significant and the corrected radiative transfer equation of Eq.~\eqref{eq:lrt_mb}, should operate. Eq.~\eqref{eq:lrt_mb} reduces to a renormalized version, $\kappa_\nu,\epsilon_\nu\!\mapsto\!f(\tau_{\nu_0}^{\mathrm{coh}})\,\kappa_\nu,\,f(\tau_{\nu_0}^{\mathrm{coh}})\,\epsilon_\nu$, of the regular radiative transfer equation, when turbulent broadening is significant over many coherence lengths. The intrinsic double-peaked core of $\phi'_{\nu}$ at large $\tau_{\nu_0}^{\mathrm{coh}}$ is then convolved away by kinematics, leaving a net suppression of effective optical depth and emissivity relative to the usual treatment of line radiative transfer. Practically, this suppression biases optical depth and column density inferences $\sim \mathcal{O}(1/\tau_{\nu_0}^{\mathrm{coh}}) \sim 1/10$, if many-body effects are ignored. 

\subsubsection{Hydrogen 21 cm line}
We now turn to the neutral hydrogen 21 cm line that originates from the magnetic dipole transition between the hyperfine states of the ground electronic level of the hydrogen atom. This line is of particular interest because of its ubiquity throughout the interstellar medium. In contrast to molecular lines, where radiative relaxation and level structure introduce complexity, the 21 cm line is in very good approximation represented by a simple two-level system, and in the ISM its level populations are typically governed by collisions alone.

We begin by computing the ratio between many-body and single-body corrections. In excellent approximation, we can set the abundance of HI to $\sim 1$, and use $h\nu/k_BT$ for the relative population difference. Assuming a thermalized radiation field $\bar{n} = k_BT/h\nu$, using $m_{\mathrm{H}}$ for the hydrogen mass, and evaluating the coherent optical depth at line center, we obtain
\begin{align}    
\frac{\tau^{\mathrm{coh}}_{\nu_0}}{A_{0} (\bar{n}+1) \tau_M/2} &\approx  \frac{3 \lambda}{4\pi}  n \frac{h^2 c^3}{(k_BT)^2} \sqrt{\frac{m_{\mathrm{H}}}{2\pi k_BT}}\nonumber \\ 
&= 7.7 \times 10^4 \left(\frac{n}{100\ \mathrm{cm}^{-3}}\right) \left(\frac{T}{50 \ \mathrm{K}} \right)^{-5/2},
\end{align}
to verify that for HI 21 cm emission or absorption from the CNM phase, it is comfortably ensured that many-body interactions dominate the radiative transfer equation corrections. 

We proceed to compute the coherent optical depth. We recall Eq.~(\ref{eq:tau_coh_app}). Again, we use the LTE approximation for the energy level difference. We assume that decorrelation proceeds through collisions only and use Eq.~(\ref{eq:tau_M_col}). We substitute the line profile for Eq.~(\ref{eq:phinu_app}). We then find the coherent optical depth at line center to be, 
\begin{align}
\tau_{\nu_0}^{\mathrm{coh}} &= \frac{1}{1+b}\frac{\lambda^2}{32 \pi} g_u A_{ul} \frac{h c^2 m\mathrm{_H}}{(k_BT)^2 \sigma}, 
\end{align}
where we recall that the factor $b$ describes the relative turbulent broadening over a coherence length and $m_\mathrm{H}$ is the hydrogen mass. Using $A_{ul} = 2.88 \times 10^{-15}$ s$^{-1}$, $g_u=3$, $\lambda=21$ cm, and adopting a typical interstellar temperature of the cold neutral medium (CNM), we find
\begin{align}
\tau_{\nu_0}^{\mathrm{coh}} &\approx 7.8\ \frac{1}{1+b} \left(\frac{T}{50\ \mathrm{K}}\right)^{-2} \left(\frac{\sigma}{10^{-15} \ \mathrm{cm}^{2}} \right)^{-1}. 
\end{align}
With the non-thermal broadening a function of the coherence length, it is interesting to put a number on, 
$\ell_{\mathrm{coh}} = 8.24\times 10^{18}\ \mathrm{cm} \ \left(\frac{n}{100\ \mathrm{cm}^{-3}} \right)^{-1} \left(\frac{T}{50 \ \mathrm{K}} \right)^{-1/2} \left(\frac{\sigma}{10^{-15}\ \mathrm{cm}^{2}} \right)^{-1},$
which, as might have been anticipated, shows that enormous path lengths of HI gas are correlated to produce many-body effects. 

The interplay between coherence effects and radiative transfer for HI 21 cm emission differs somewhat from that discussed for CO line emission, in that for CO, $\tau_{\nu} \gg \tau_{\nu}^{\mathrm{coh}}$, while for the HI 21 cm line, $\tau_{\nu} \sim \tau_{\nu}^{\mathrm{coh}}$. Although HI 21 cm traces the most abundant constituent of the ISM, its magnetic dipole nature renders the transition intrinsically weak. Most HI emission is therefore optically thin, but moderate optical depths can be exhibited by features or clouds of the CNM phase of the ISM. We cite here the HI optical depth of a typical CNM cloud,
\begin{align}
\tau_{\nu}^{\mathrm{cloud}} &= 2.7\ \frac{1}{1 + \mathcal{M}_{\mathrm{cloud}}} \left[\frac{n}{100\ \mathrm{cm}^{-3}} \right] \nonumber \\ &\times\left[\frac{\ell_{\mathrm{cloud}}}{10^{18} \ \mathrm{cm}}\right]  \left[\frac{T}{50 \ \mathrm{K}} \right]^{-3/2},
\end{align}
where $\ell_{\mathrm{cloud}}$ is the cloud length along the line-of-sight and $\mathcal{M}_{\mathrm{cloud}}$ the cloud Mach number that we include to represent the non-thermal broadening. With the coherence length and optical depth, and the cloud length and optical depth comparable, it is sometimes, if not usually for HI, the cloud dimensions that limit the number of atoms that can become correlated to produce the many-body effects that impact the transfer of radiation. In such cases, the coherent optical depth approaches the cloud optical depth. We have discussed the implications for the (many-body) radiative transfer of such lines in Sec.~\ref{sec:sol_rt}. We found that for $\tau_{\nu}^{\mathrm{coh}}=\tau_{\nu}>1$, the emission yields of such lines plateau at $1-e^{-1}$ times the source function. Conversely, absorption for such lines will not completely diminish the background source intensity, but absorption will saturate at intensity levels $1/e$ times the background source intensity, regardless of the optical depth. We should therefore expect that many-body effects systematically bias inferences of both the HI spin temperature and optical depth. Likely leading to overestimates of the spin temperature, and underestimates of the `real' optical depth. 
\subsubsection{Impact of model assumptions and radiative transfer}
The results presented here are based on a model that assumes dominant Doppler broadening, equal lifetimes for the transition levels, and isotropic, Boltzmann-distributed particle positions and velocities. These simplifications have been sufficient to demonstrate, in a general sense, how many-body effects modify radiative transfer, and to show that in interstellar gases the low-frequency transitions of cold CO and HI are particularly susceptible. The framework can be refined by relaxing these assumptions, for example by implementing a first-principles treatment of level-specific lifetimes, or by incorporating spatially varying excitation and velocity fields in a realistic radiative transfer calculation tailored to specific astrophysical environments.

\section{Conclusions}
In this work, we have derived a generalized radiative transfer equation that explicitly accounts for many-body interactions between a radiation field and an ensemble of molecules. Starting from first principles, we showed that the canonical radiative transfer equation is recovered only in the limit of low coherent optical depth, while for $\tau_\nu^{\mathrm{coh}} \gtrsim 1$ higher-order collective effects modify the effective absorption and emission terms by a well-defined multiplicative factor. To our knowledge, this represents the first derivation of the radiative transfer equation that rigorously incorporates many-body correlations into line transport theory.

We have identified a clear physical regime in which these many-body effects should manifest, set by the interplay between molecular lifetimes, Doppler decorrelation, collisional dephasing, and optical depth. Within this regime, we have proposed a room temperature laboratory experiment, using a low-pressure, moderately optically thick transition, that should reveal the predicted suppression of radiative coupling due to many-body effects with relative ease. Such a measurement would not only provide a direct demonstration of interparticle correlations in a thermal gas, but could also serve as a practical method for determining collisional broadening and lifetime coefficients.

Applying our framework to astrophysical conditions, we find that many-body effects should naturally occur in two of the most important tracers of the interstellar medium: the HI 21\,cm line and the low-$J$ rotational transitions of CO. Under cold ISM conditions, both transitions can reach coherent optical depths of order unity or higher, implying that many-body corrections should alter emergent line profiles and bias standard inferences of gas properties. Accounting for these effects may be essential for accurately interpreting emission from these principal tracers of dense and diffuse interstellar gas.

\begin{acknowledgments}
Support for this work from the Swedish Research Council (VR) under grant number 2021-00339 is acknowledged. I am grateful to Martin Houde, for insightful discussions and encouragement throughout the preparation of this manuscript. 
\end{acknowledgments}

\bibliographystyle{apsrev4-2}
\bibliography{lib}  
\newpage
\onecolumngrid
\appendix


\section{Decorrelation of the particle operators}
In this section, we evaluate the expectation value of a series of particle operators, that include time-dependence through collisional coupling to a bath. We work out a general term, that occurs when evaluating the many-body expansion elements of the stimulated radiative operator. The $n$'th order term, is associated with the following general series of particle operators, that we trace over the particle density operator and the bath,
\begin{align}
\label{eq:gen_time_decay}
\mathrm{tr} \left\{ \tilde{R}_{j_1+}(t_1) \tilde{R}_{j_1-}(t_2)  \tilde{R}_{j_2+}(t_3) \tilde{R}_{j_2-}(t_4) \cdots  \tilde{R}_{j_{n+1}+}(t_{2n+1}) \tilde{R}_{j_{n+1}-}(t_{2n+2})  \hat{\rho}_0^{(M)} \otimes \hat{\rho}_0^{(B)}\right\}.
\end{align}
We recall that $\tilde{R}_{j\pm} (t)= \hat{U}^{\dagger}(t) \hat{R}_{j\pm} \hat{U}(t)$, and we use the general relation for a time-evolution operator $ \hat{U}(t) \hat{U}^{\dagger}(t') = \hat{U}(t-t')$. Using this relation, we point out that we can factorize $\hat{U}(t_1) = \hat{U}(t_{1} - t_{2}) \hat{U}(t_{2} - t_{3}) \cdots  \hat{U}(t_{2n-1} - t_{2n})  \hat{U}(t_{2n})$, whilst noting that by the integral limits, it is ensured that $t_1 > t_2 > t_3 > \cdots$. Using these relations, we can rearrange Eq.~(\ref{eq:gen_time_decay}) to
\begin{align}
\label{eq:time_trace}
\mathrm{tr}& \left\{\hat{U}^{\dagger}(t_{2n+2}) \hat{U}^{\dagger}(t_{2n+1} - t_{2n+2})  \hat{U}^{\dagger}(t_{2n} - t_{2n+1}) \cdots  \hat{U}^{\dagger}(t_{1} - t_{2}) \right. \nonumber \\
&\times \left. \hat{R}_{j_1+} \hat{U}(t_1-t_2) \hat{R}_{j_1-} \hat{U}(t_2-t_3)  \hat{R}_{j_2+} \hat{U} (t_3-t_4) \hat{R}_{j_2-} \cdots  \hat{R}_{j_{n+1}+} \hat{U}(t_{2n+1} - t_{2n+2}) \tilde{R}_{j_{n+1}-} \hat{U} (t_{2n+2}) \hat{\rho}_0^{(M)} \otimes \hat{\rho}_0^{(B)}\right\}.
\end{align}
We proceed by averaging the particle-perturber collisional propagation operators over all perturber configurations by tracing out $\hat{\rho}^{(P)}(0)$, so that, $\mathrm{tr} \{ \hat{U}(t) \hat{\rho}_0^{(B)} \} \to \hat{A}(t)$, and the time-propagation operators lose their unitary properties. In addition, we use the cyclic property of the trace, and assume the sample to be equilibrated before start of the first interaction at $t_{2n+2}$: $\hat{A} (t_{2n+2}) \hat{\rho}_0^{(M)} \hat{A}^{\dagger} (t_{2n+2}) = \hat{\rho}_0^{(M)}$. We use that at $t=0$, the molecules in the ensemble are uncorrelated: $\hat{\rho}_0^{(M)} = \bigotimes_j \hat{\rho}_0^{(j,M)}$, and the evolution operator can be decomposed into sub-operators working on a specific molecule:
$$
\hat{A}(t) = \hat{A}_1 (t) \otimes \hat{1}_2 \otimes \cdots \otimes \hat{1}_N + \hat{1}_1 \otimes \hat{A}_2 (t) \otimes \cdots \otimes \hat{1}_N + \cdots + \hat{1}_1  \otimes \hat{1}_2 \otimes \cdots \otimes \hat{A}_N (t),
$$ 
where $\hat{A}_j(t)$ is an operator only working on molecule $j$. Using this, and noting $\tau_i = t_{i+1} - t_i$, we can factorize the trace of Eq.~(\ref{eq:time_trace}) into components specific to individual molecular terms
\begin{align}
\label{eq:time_trace_2}
&\mathrm{tr} \left\{\hat{A}^{\dagger}(\tau_{2n+1})  \hat{A}^{\dagger}(\tau_{2n}) \cdots  \hat{A}^{\dagger}(\tau_{1}) \hat{R}_{j_1 +} \hat{A}(\tau_1)\hat{R}_{j_1-} \hat{A} (\tau_2) \hat{R}_{j_2 +} \hat{A}(\tau_3) \cdots \hat{R}_{j_{n+1} -} \hat{\rho}_0^{(M)}  \right\} \nonumber \\
&= \mathrm{tr} \left\{\hat{A}^{\dagger}(\tau_{2n+1})  \hat{A}^{\dagger}(\tau_{2n}) \cdots  \hat{A}^{\dagger}(\tau_{1}) \hat{R}_{j_1 +} \hat{A}(\tau_1)\hat{R}_{j_1-} \hat{A} (\tau_2) \hat{A}(\tau_3) \cdots \hat{A} (\tau_{2n+1})\hat{\rho}_0^{(j_1,M)}  \right\} \nonumber \\
&\times  \mathrm{tr} \left\{\hat{A}^{\dagger}(\tau_{2n+1})  \hat{A}^{\dagger}(\tau_{2n}) \cdots  \hat{A}^{\dagger}(\tau_{3}) \hat{R}_{j_2 +} \hat{A}(\tau_3)\hat{R}_{j_2-} \hat{A} (\tau_4) \hat{A}(\tau_5) \cdots \hat{A} (\tau_{2n+1})\hat{\rho}_0^{(j_2,M)}  \right\} \nonumber \\
&\times  \mathrm{tr} \left\{\hat{A}^{\dagger}(\tau_{2n+1})  \hat{A}^{\dagger}(\tau_{2n}) \cdots  \hat{A}^{\dagger}(\tau_{3}) \hat{R}_{j_3 +} \hat{A}(\tau_5)\hat{R}_{j_2-} \hat{A} (\tau_6) \hat{A}(\tau_7) \cdots \hat{A} (\tau_{2n+1})\hat{\rho}_0^{(j_3,M)} \right\} \nonumber \\
&\vdots \nonumber \\
&\times  \mathrm{tr} \left\{\hat{A}^{\dagger}(\tau_{2n+1})  \hat{A}^{\dagger}(\tau_{2n}) \hat{A}^{\dagger}(\tau_{2n-1}) \hat{R}_{j_{n} +} \hat{A}(\tau_{2n-1})\hat{R}_{j_{n}-} \hat{A} (\tau_{2n}) \hat{A}(\tau_{2n+1}) \hat{\rho}_0^{(j_n,M)}  \right\} \nonumber \\
&\times  \mathrm{tr} \left\{\hat{A}^{\dagger}(\tau_{2n+1}) \hat{R}_{j_{n+1} +} \hat{A}(\tau_{2n+1})\hat{R}_{j_{n+1}-} \hat{\rho}^{(M)}_{j_{n+1}} (0)  \right\} .
\end{align}
Then using Eq.~(\ref{eq:A_matrixelements}), together with $\mathrm{tr}\{ \hat{R}_{j+}\hat{R}_{j-} \hat{\rho}_0^{(j,M)} \} = n_+$, allows us to evaluate the time-dependencies in Eq.~(\ref{eq:time_trace_2}), and evaluate the traces, to obtain the result 
\begin{align}
\label{eq:collision_decor}
\mathrm{tr}& \left\{ \tilde{R}_{j_1+}(t_1) \tilde{R}_{j_1-}(t_2)  \tilde{R}_{j_2+}(t_3) \tilde{R}_{j_2-}(t_4) \cdots  \tilde{R}_{j_{n+1}+}(t_{2n+1}) \tilde{R}_{j_{n+1}-}(t_{2n+2})  \hat{\rho}_0^{(M)} \right\} \nonumber \\ &= n_+^{n+1} \left(\prod_{i=1}^{n+1} e^{-\tau_{2i-1}[\tau_L^{-1} + (i-1)\tau_{M}^{-1}]} \right) \left( \prod_{i=1}^{n} e^{-\tau_{2i} \frac{i}{\tau_{M}}}
\right) ,
\end{align}
where we parsed the products of decorrelation functions into those that depend on even and uneven $i$ in the time-variables $\tau_i$. In this manuscript, we have been interested in lines that are dominantly Doppler broadened. Doppler broadening manifests through a decorrelation function, $e^{-\tau_i^2/\tau_D^2}$, at uneven $i$. For the expansion of the stimulated radiative operator, with $\tau_D \ll \tau_M,\tau_L$, decorrelations captured in $\tau_L$ and $\tau_M$ can in fact be neglected during $\tau_i$ with uneven $i$. In contrast, during times $\tau_i$, of even $i$, the dominant source of decorrelation is captured in $\tau_M$. Finally, we can work out explicitly, the time-integrals 
\begin{align}
\int d\tau_2 \int d\tau_4 \int d\tau_6 \cdots \int d\tau_{2n} \left( \prod_{i=1}^{n} e^{-\tau_{2i} \frac{i}{\tau_{M}}}
\right) = \frac{n!}{\tau_M^{n}},
\end{align}
that emerge when evaluating the $n$'th order many-body correction to the stimulated radiative operator.

\section{Forcing the forward direction for many-body radiative transfer corrections}
In formulating the radiative operators $\hat \sigma (t,t')$, and $\hat \gamma (t,t')$, we put the factor $e^{i(\boldsymbol{k}-\boldsymbol{k}')\cdot \hat{\boldsymbol{r}}_j} \to \delta_{\boldsymbol{k},\boldsymbol{k}'}$, claiming that this was allowed when $L\gg \lambda$, and because subsequent many-body correction terms that originate from the time-evolution of the density operator do not interfere with this term. Here we validate this approach when considering radiative transfer in the regime where many-body effects dominate. We call the stimulated radiative operator, that does not make the simplification $e^{i(\boldsymbol{k}-\boldsymbol{k}')\cdot \hat{\boldsymbol{r}}_j} \to \delta_{\boldsymbol{k},\boldsymbol{k}'}$, $\hat{\Sigma}(t,t')$, and recall Eq.~(\ref{eq:rt_double_exp}), to note it
\begin{align}
\hat{\Sigma}(t,t') = \sum_{j \boldsymbol{k}'} g_{\boldsymbol{k}} g_{\boldsymbol{k}'} [\tilde{R}_{j+}(t),\tilde{R}_{j-}(t')]\left[\frac{2 h\nu}{\lambda^2} \hat{a}_{\boldsymbol{k}'}^{\dagger} \hat{a}_{\boldsymbol{k}} \right] \ e^{i(\omega_0-\omega_k)t} e^{-i(\omega_0-\omega_{k'})t'} e^{i \boldsymbol{k} \cdot \hat{\boldsymbol{v}}_j t} e^{-i \boldsymbol{k}' \cdot \hat{\boldsymbol{v}}_j t'} e^{i (\boldsymbol{k}-\boldsymbol{k}') \cdot \hat{\boldsymbol{r}}_j} .
\end{align}
The next-order correction to the stimulated radiative operator is
\begin{align}
\hat{\Sigma}(t,t') \to -\hbar^{-2} \left[\left[\hat{\Sigma}(t,t'),\hat{V}(t'')\right],\hat{V}(t''')\right] ,
\end{align}
where in Eq.~(\ref{eq:kv_1_comm_parsed}), we had shown that this correction can be parsed into many-body and single-body terms. We specialize to the regime where many-body corrections dominate. The many-body correction to the (proper) stimulated radiative operator is then
\begin{align}
\label{eq:sigma_large_pert}
-\hbar^{-2} \sum_{j \boldsymbol{k}'} \frac{g_{\boldsymbol{k}} g_{\boldsymbol{k}'}}{c} [\tilde{R}_{j+}(t),\tilde{R}_{j-}(t')] \ e^{i(\omega_0-\omega_k)t} e^{-i(\omega_0-\omega_{k'})t'} e^{i \boldsymbol{k} \cdot \hat{\boldsymbol{v}}_j t} e^{-i \boldsymbol{k}' \cdot \hat{\boldsymbol{v}}_j t'} e^{i (\boldsymbol{k}-\boldsymbol{k}') \cdot \hat{\boldsymbol{r}}_j} \left[\left[\frac{2 h\nu}{\lambda^2} \hat{a}_{\boldsymbol{k}'}^{\dagger} \hat{a}_{\boldsymbol{k}},\hat{V}(t'')\right],\hat{V}(t''')\right] .
\end{align}
We parse out the commutator, $\left[\left[\frac{2 h\nu}{\lambda^2} \hat{a}_{\boldsymbol{k}'}^{\dagger} \hat{a}_{\boldsymbol{k}},\hat{V}(t'')\right],\hat{V}(t''')\right] = \frac{2 h\nu}{\lambda^2}\left[\left[ \hat{a}_{\boldsymbol{k}'}^{\dagger} \hat{a}_{\boldsymbol{k}},\hat{B}(t'')\right],\hat{B}^{\dagger}(t''')\right] + \mathrm{c.c.},$ and work out
\begin{align}
&\left[\left[ \hat{a}_{\boldsymbol{k}'}^{\dagger} \hat{a}_{\boldsymbol{k}},\hat{B}(t'')\right],\hat{B}^{\dagger}(t''')\right] = \sum_{lm\ \boldsymbol{k}''\boldsymbol{k}'''} g_{\boldsymbol{k}''} g_{\boldsymbol{k}'''} \left[\left[ \hat{a}_{\boldsymbol{k}'}^{\dagger} \hat{a}_{\boldsymbol{k}},\hat{a}_{\boldsymbol{k}''} \tilde{R}_{l+}(t'')\right],\hat{a}_{\boldsymbol{k}'''}^{\dagger} \tilde{R}_{m-}(t''')\right] \nonumber \\ &\times
e^{i(\omega_0-\omega_{k''})t''} e^{-i(\omega_0-\omega_{k'''})t'''} e^{i \boldsymbol{k}''\cdot \hat{\boldsymbol{r}}_{l}}e^{i \boldsymbol{k}''\cdot \hat{\boldsymbol{v}}_l t''} e^{-i \boldsymbol{k}'''\cdot \hat{\boldsymbol{r}}_{m}} e^{-i \boldsymbol{k}'''\cdot \hat{\boldsymbol{v}}_m t'''} \nonumber \\ 
&= \sum_{lm\ \boldsymbol{k}'''} g_{\boldsymbol{k}'} g_{\boldsymbol{k}'''} \left[\hat{a}_{\boldsymbol{k}}\tilde{R}_{l+}(t''),\hat{a}_{\boldsymbol{k}'''}^{\dagger} \tilde{R}_{m-}(t''')\right] 
e^{i(\omega_0-\omega_{k'})t''} e^{-i(\omega_0-\omega_{k'''})t'''} e^{i \boldsymbol{k}'\cdot \hat{\boldsymbol{r}}_{l}}e^{i \boldsymbol{k}'\cdot \hat{\boldsymbol{v}}_l t''} e^{-i \boldsymbol{k}'''\cdot \hat{\boldsymbol{r}}_{m}} e^{-i \boldsymbol{k}'''\cdot \hat{\boldsymbol{v}}_m t'''} \nonumber \\ 
&= \sum_{l \ \boldsymbol{k}'''}  g_{\boldsymbol{k}'} g_{\boldsymbol{k}'''} \hat{a}_{\boldsymbol{k}'''}^{\dagger} \hat{a}_{\boldsymbol{k}} \left[\tilde{R}_{l+}(t''), \tilde{R}_{l-}(t''')\right] e^{i(\omega_0-\omega_{k'})t''} e^{-i(\omega_0-\omega_{k'''})t'''} e^{i (\boldsymbol{k}'-\boldsymbol{k}''')\cdot \hat{\boldsymbol{r}}_{l}} e^{i \boldsymbol{k}' \cdot \hat{\boldsymbol{v}}_l t''} e^{-i \boldsymbol{k}'''\cdot \hat{\boldsymbol{v}}_m t'''} \nonumber \\ &+
\sum_{lm} g_{\boldsymbol{k}'} g_{\boldsymbol{k}} \tilde{R}_{l+}(t'')\tilde{R}_{m-}(t''')
e^{i(\omega_0-\omega_{k'})t''} e^{-i(\omega_0-\omega_{k})t'''} e^{i \boldsymbol{k}'\cdot \hat{\boldsymbol{r}}_{l}}e^{i \boldsymbol{k}'\cdot \hat{\boldsymbol{v}}_l t''} e^{-i \boldsymbol{k}\cdot \hat{\boldsymbol{r}}_{m}} e^{-i \boldsymbol{k}\cdot \hat{\boldsymbol{v}}_m t'''}.
\end{align}
We focus on the first term in the above sum, that bears a close resemblance to the initial $\hat{\Sigma}(t,t')$ operator. We rewrite the phase function $e^{i (\boldsymbol{k}'-\boldsymbol{k}''')\cdot \hat{\boldsymbol{r}}_{l}} = e^{i (\boldsymbol{k}'-\boldsymbol{k})\cdot \hat{\boldsymbol{r}}_{l}} e^{i (\boldsymbol{k}-\boldsymbol{k}''')\cdot \hat{\boldsymbol{r}}_{l}}$, and absorb the phase $e^{i (\boldsymbol{k}-\boldsymbol{k}')\cdot \hat{\boldsymbol{r}}_{j}}$ from Eq.~(\ref{eq:sigma_large_pert}) to isolate the element $e^{i (\boldsymbol{k}-\boldsymbol{k}')\cdot (\hat{\boldsymbol{r}}_{j}-\hat{\boldsymbol{r}}_{l})}$. At separations much larger than the wavelength, the phase between particles $j$ and $l$ forces the forward direction, and reduces to $\delta_{\boldsymbol{k},\boldsymbol{k'}}$. However, there can be a significant amount of particles within an order of the wavelength, for which this forcing doesn't apply. To an order of magnitude, we can divide the sum over particles, over those particles within a wavelength sphere, that can act coherently $\sim \mathcal{N} \lambda^3$, and for which $e^{i (\boldsymbol{k}-\boldsymbol{k}')\cdot (\hat{\boldsymbol{r}}_{j}-\hat{\boldsymbol{r}}_{l})} \approx 1$, and all other particles that are farther away, and for which $e^{i (\boldsymbol{k}-\boldsymbol{k}')\cdot (\hat{\boldsymbol{r}}_{j}-\hat{\boldsymbol{r}}_{l})} \approx \delta_{\boldsymbol{k},\boldsymbol{k}'}$. It can be shown that the number of coherently interacting particles is confined to those particles that are resonant and are of order $N_{\mathrm{coh}}\sim \frac{A_0}{\Delta \omega} \mathcal{N}\lambda^3$, while the number of particles that interact through the many-body effect studied in this paper has been shown to be $N_{\mathrm{m.b.}}\sim \frac{A_0}{\Delta \omega} \mathcal{N} \lambda^2 \ell_{\mathrm{coh}}\sim \tau_{\nu}^{\mathrm{coh}}$. Their relative magnitudes are $$\frac{\ell_{\mathrm{coh}}}{\lambda} \sim \frac{c
}{v_{\mathrm{th}} \sigma \lambda n} \sim 10^{6} \left(\frac{\lambda}{\mathrm{cm}}\right)^{-1} \left(\frac{n}{10^{15}\ \mathrm{cm}^{-3}}\right)^{-1} \left(\frac{\sigma}{10^{-15}\ \mathrm{cm}^2}\right)^{-1} \left(\frac{T}{296 \ \mathrm{K}}\right)^{-1/2} \left(\frac{M}{60 \ \mathrm{m_H}} \right)^{1/2},$$
showing clear dominance by many-body effects for practically all Doppler broadened lines. With $N_{\mathrm{m.b.}}/N_{\mathrm{coh}}\gg 1$, we may invoke the forcing of the forward direction in the phase $e^{i (\boldsymbol{k}-\boldsymbol{k}')\cdot (\hat{\boldsymbol{r}}_{j}-\hat{\boldsymbol{r}}_{l})} \to \delta_{\boldsymbol{k},\boldsymbol{k}'}$ in the large sample limit $L \sim \ell_{\mathrm{coh}} \gg \lambda$. We recognize then, that the many-body correction to the stimulated radiative operator,
\begin{align}
\hat{\Sigma} (t,t') &\to -\hbar^{-2} \left[\left[\hat{\Sigma}(t,t'),\hat{V}(t'')\right],\hat{V}(t''')\right] \nonumber \\
&= \left[\sum_j2 g_{\boldsymbol{k}}^2 \tilde{R}_{j3}(t,t') f_{j}^{\boldsymbol{k}}(t-t') \right] \left[\Sigma(t'',t''') + \Gamma(t'',t''') \right],
\end{align}
is analogous to Eq.~(\ref{eq:sig_corr}), and where 
$$\hat{\Gamma}(t,t') = \sum_{jl} \frac{2 h\nu}{\lambda^2} g_{\boldsymbol{k}}^2  \tilde{R}_{j+}(t) \tilde{R}_{l-}(t') e^{i(\omega_0-\omega_k)(t-t')} e^{i \boldsymbol{k} \cdot \hat{\boldsymbol{v}}_jt} e^{-i \boldsymbol{k} \cdot \hat{\boldsymbol{v}}_lt'} e^{i \boldsymbol{k} \cdot (\hat{\boldsymbol{r}}_j - \hat{\boldsymbol{r}}_l)},
$$
is the (proper) spontaneous radiative operator. Finally, we note that $\mathrm{tr}\left\{ \hat \Sigma(t,t') \hat{\rho}_0\right\} = \mathrm{tr}\left\{ \hat \sigma(t,t') \hat{\rho}_0\right\}$ and $\mathrm{tr}\left\{ \hat \Gamma(t,t') \hat{\rho}_0\right\} = \mathrm{tr}\left\{ \hat \gamma(t,t') \hat{\rho}_0\right\}$. This validates the forcing of the forward direction, when performing many-body corrections.

\end{document}